\documentclass[12pt]{article}
\usepackage[body={17cm, 23cm, centered}]{geometry}
\usepackage{epsfig,verbatim,cite}
\usepackage{braket}
\usepackage{lmodern}

\usepackage[utf8]{inputenc}

\usepackage[T1]{fontenc}
\usepackage{mathtext,marvosym,textcomp}
\usepackage{mathtools}
\usepackage{slashed}
\usepackage[usenames,dvipsnames,svgnames,table]{xcolor}
\usepackage{tikz}
\usepackage[normalem]{ulem}
\usepackage{tocloft}
\usepackage{epsfig,amsmath,amsfonts,amssymb,arydshln}
\usepackage[makeroom]{cancel}
\usepackage{multirow}

 \usepackage[debug,pageanchor=false]{hyperref}
\hypersetup{colorlinks=true,linktocpage,breaklinks,
            urlcolor=blue,
            linkcolor=red,
            citecolor=blue
            }

\numberwithin{equation}{section}

\def\a{\alpha} \def\b{\beta} \def\g{\gamma} \def\d{\delta} \def\e{\epsilon}
\def\ve{\varepsilon}  \def\h{\eta} \def\q{\theta}
  \def\k{\kappa} \def\l{\lambda} \def\m{\mu}
\def\n{\nu} \def\x{\xi}   \def\r{\rho}
 \def\s{\sigma} \def\t{\tau}  \def\f{\varphi}

\def\D{\Delta} 
   \def\Q{\Theta}
   \def\L{\Lambda} 
 \def\X{\Xi}  
   
\def\F{\Phi}  \def\Y{\Psi} \def\W{\Omega}

\def\ba{\bar{a}}\def\bb{{\bar{b}}}\def\bc{\bar{c}}\def\bd{\bar{d}}
\def\bl{\bar{l}}\def\bm{\bar{m}}\def\bn{\bar{n}}
\def\bs{\bar{s}}

\def\fr{\frac}  \def\dt{\partial}

\def\ph{\phantom}
\def\mc{\mathcal}
\def\mA{\mathcal{A}}
\def\mB{\mathcal{B}}
\def\mC{\mathcal{C}}
\def\mD{\mathcal{D}}
\def\mE{\mathcal{E}}
\def\mF{\mathcal{F}}
\def\mG{\mathcal{G}}
\def\mH{\mathcal{H}}
\def\mK{\mathcal{K}}
\def\mL{\mathcal{L}}
\def\mM{\mathcal{M}}
\def\mN{\mathcal{N}}
\def\mP{\mathcal{P}}
\def\mQ{\mathcal{Q}}
\def\mR{\mathcal{R}}

\def\mZ{\mathcal{Z}}
\def\mS{\mathcal{S}}

\def\hM{\hat{M}}

\def\tm{$\times$}
\def\od{$\odot$}

\def\bm{{\bar{\mu}}}
\def\bn{{\bar{\nu}}}
\def\bl{{\bar{\lambda}}}
\def\br{{\bar{\rho}}}
\def\bs{{\bar{\sigma}}}

\def\hr{\hat{\rho}}

\def\XX{\mathbb{X}}

\newcommand\bqa {\begin{eqnarray}}
\newcommand\eqa {\end{eqnarray}}

\newcommand{\bear}{\begin{array}}
\newcommand{\enar}{\end{array}}

\newcommand{\ol}[1]{\overline{#1}}

\newcommand{\be}{\begin{equation}}
\newcommand{\ee}{\end{equation}}
\def\bea{\begin{eqnarray}}
\def\eea{\end{eqnarray}}

\def\M{{\mathcal{M}}}
\def\DD{{\mathcal{D}}}

\newcommand{\hmu}{{\hat{\mu}}}
\newcommand{\hnu}{{\hat{\nu}}}

\newcommand{\hrh}{\hat{\rho}}

\newcommand{\hsg}{\hat{\sigma}}
\newcommand{\hmH}{\hat{\mH}}

\allowdisplaybreaks

\begin{document}
\renewcommand{\contentsname}{}
\renewcommand{\refname}{\begin{center}References\end{center}}
\renewcommand{\abstractname}{\begin{center}\footnotesize{\bf Abstract}\end{center}} 

\begin{titlepage}
\ph{preprint}

\vfill

\begin{center}
   \baselineskip=16pt
   {\large \bf    Exotic potentials and Bianchi identities\\in SL(5) exceptional field theory.    
   }
   \vskip 2cm
    Kirill Gubarev$^\dagger$\footnote{\tt kirill.gubarev@phystech.edu }
    Edvard T. Musaev$^\dagger{}^\bullet$\footnote{\tt musaev.et@phystech.edu}
       \vskip .6cm
             \begin{small}
                          {\it
                          $^\dagger$Moscow Institute of Physics and Technology, 
                          Laboratory of High Energy Physics\\
                          Institutskii per., 9, 141702, Dolgoprudny, Russia\\[0.5cm] 
                          $^\bullet$Kazan Federal University, Institute of Physics\\
                          Kremlevskaya 16a, 420111, Kazan, Russia                         
                          } \\ 
\end{small}
\end{center}

\vfill 
\begin{center} 
\textbf{Abstract}
\end{center} 
\begin{quote}
Tensor hierarchy of Exceptional Field Theories contains gauge fields satisfying certain Bianchi identities. We define the full set of fluxes of the SL(5) exceptional field theory containing known gauge field strengths, generalized anholonomy coefficients and two new fluxes. It is shown that the full SL(5) ExFT Lagrangian can be written in terms of the listed fluxes. We derive the complete set of Bianchi identities and identify magnetic potentials of the theory and the corresponding (wrapped) membranes of M-theory.
\end{quote} 
\vfill
\setcounter{footnote}{0}
\end{titlepage}

\tableofcontents

\setcounter{page}{2}

\section{Introduction}

Theories of fundamental extended objects pretending on a quantum description of gravity, be it string theory or M-theory, are essentially higher dimensional. Hence one is naturally interested in dimensional reductions in order to reproduce effectively four dimensional physics. To have phenomenologically feasible dynamics it is typically necessary to go beyond simple toroidal compactifications and to consider models that include branes wrapping topological cycles of a compact manifold, generating non-trivial fluxes along them, as well as orientifold planes to satisfy tadpole cancellation conditions (see e.g. \cite{Grana:2005jc} for a review). Fluxes being a natural component of any string compactification scheme in addition to supersymmetry breaking generate non-zero masses for scalar fields determining the low energy effective potential. For example in the case of Type IIB compactifications on Calabi-Yau three-folds with O3 planes the so-called Gukov-Vafa-Witten superpotential \cite{Gukov:1999ya}
\begin{equation}
    \label{eq:GVW}
    W = \int_{CY_3} \W \wedge  (F_3 + i \t H_3),
\end{equation}
where $F_3$ and $H_3$ are 3-form fluxes from the R-R and NS-NS sectors respectively and $\t = C_0 + i e^{-\f}$ is the axio-dilaton. $\W$ is the Calabi-Yau $(3,0)$-form defining its topology. 

The tadpole cancellation conditions for branes interacting with the fluxes magnetically from the point of view of the low energy effective theory (supergravity) can be formulated in terms of Bianchi identities. Schematically these read
\begin{equation}
    dF + F\wedge F = \sum Q,
\end{equation}
where on the RHS we have a sum of the corresponding charges from all the sources. Tadpole cancellation condition requires that for the compact manifold the overall sum of all charges must be zero, that in particular requires adding Op-planes in addition to Dp-branes to compensate their RR charges. Using internal symmetries of string theory, T- and S-dualities, one figures out additional terms to be added to tadpoles coming from duality partners of Dp-branes, that are NS branes and exotic branes.  For example, the GVW superpotential is explicitly symmetric under  the S-duality symmetry of Type IIB theory and can be generated from the first term solely. The first term includes the 3-form field strength generated by a D5-brane wrapping the internal 3-cycle, while the second term comes from its S-dual NS5-brane. Putting this logic forward one includes T-duality and more generally U-duality into the chain, generating more terms in the superpotential \cite{Lombardo:2016swq}. Such fluxes generated by duality transformations are sourced by exotic branes, in addition each of the fluxes included into a compactification scheme must fulfill tadpole cancellation conditions in the form of Bianchi identities.

When speaking purely about compactifications of 10/11-dimensional supergravity when fluxes are understood as integrals along topological cycles, the picture is significantly simplified when the formalism of embedding tensor is used \cite{deWit:2002vt,deWit:2004yr,deWit:2007mt,deWit:2008ta,Samtleben:2008pe}. In this approach fluxes related by (U-)duality transformations are various components of a spurious object called embedding tensor , that encodes the most general deformation of a $D$-dimensional (half-)maximal supergravity by a set of constant parameters, preserving supersymmetry. The embedding tensor transforms in a linear representation of the corresponding U-duality group $E_{d(d)}$, where $d=11-D$, and satisfies a set of constraints, required by the maximal supersymmetry and invariance of the theory under local transformations. The linear constraint simply projects the embedding tensor to certain irreps of $E_{d(d)}$, while the quadratic constraint can be schematically written as
\begin{equation}
    \mathbb{P}\Q\Q = 0.
\end{equation}
Here $\mathbb{P}$ is a projector to certain irreps of the U-duality group, that depends on the group chosen and whose explicit form will be given later for the case $d=4$. The embedding tensor is denoted by $\Q$ suppressing all indices. Written in terms of the geometric $GL(d)$ subgroup the above is nothing but the full collection of tadpole cancellation conditions for all fluxes along the compact $d$-dimensional space. Including those, sources by exotic branes. 

In democratic formulation of Type II supergravity Bianchi identities are simply another way of writing field equations of the theory, that appears to be more convenient when dynamics of magnetically charged objects is taken into account. As the simplest example consider the fundamental Type II string interacting with the Kalb-Ramond 2-form $B_2$ whose field strength we denote by $H_3 =d B_2$. Its field equation read
\begin{equation}
    *\, d\big(e^{-2\f}* H_3 + \dots \big)= j^{F1}_2,
\end{equation}
where $*$ denotes Hodge star in 10d and ellipses denote contributions from RR fields. Current of the fundamental string is denoted by $j_2^{F1}$ and is proportional to a two-form supported on the string world-sheet. Bianchi identities for the 3-form flux are simply $dH_3 = 0$. Now, one notices that in the absence of the string current, $j_2^{F1}=0$, the field equations above at least locally can be solve by 
\begin{equation}
    e^{-2\f}*H_3 + \dots = dB_6.
\end{equation}
Introducing the magnetically dual 7-form field $H_7 = e^{-2\phi} * H_3$, we write field equations as its Bianchi identities
\begin{equation}
    dH_7 + \dots = 0.
\end{equation}
On the other hand, Bianchi identities $d H_3 = 0$ are now written in the form of field equations, that with a source in the RHS take the form
\begin{equation}
    * d (* H_7 + \dots) = j_6^{NS5}.
\end{equation}
The source is nothing but the current for the NS5-brane, the magnetic dual of the fundamental string. Hence, we see, that Bianchi identities hold information about branes magnetically charged with the corresponding flux. 

Apparently, the same is true when fluxes are combined in duality multiplets, inducing the same for currents of the corresponding branes. In this case, say, the embedding tensor must be understood as a generalized field strength rather than a constant tensor, and the quadratic constraints become
\begin{equation}
    \dt \Q + \mathbb{P}\Q\Q = \mc{J}. 
\end{equation}
To make more precise sense of the derivative in the first term and the current on the RHS one should turn to a U-duality symmetric formulation of 11D supergravity, that is provided by exceptional field theory. Field content of exceptional field theory is given by a metric, that is a weighted scalar under the U-duality group, a set of tensor gauge fields transforming linearly and a set of scalars parametrizing a coset space $G/K$, where $G$ is the U-duality group and $K$ is its maximal compact subgroup. Covariance under $G$ is achieved by extending the underlying space-time, whose points are now labeled by coordinates $(x^\m, \XX^{\mc{M}})$. Here $x^\m$ are the usual coordinates on the so-called external space-time  of dimension $D$, and $\XX^{\mc{M}}$ are the so-called extended coordinates. The index $\mc{M} = 1,\dots,n_v$ labels components of the irrep $R_V$, that is precisely the one in which vector fields $A_\m{}^{\mc{M}}$ transform. We will return to a more detailed definition of exceptional field theory in Section \ref{sec:SL5}. For the original papers on the subject the reader is referred to \cite{Berman:2010is,Berman:2011jh,Hohm:2013pua,Hohm:2013vpa,Hohm:2013uia,Abzalov:2015ega,Musaev:2015ces}, and for a review see \cite{Hohm:2019bba,Musaev:2019zcr,Berman:2020tqn}. 

Our goal in this work is to derive all Bianchi identities of the SL(5) exceptional field theory. For that we define a set of quantities that can be referred to as fluxes in addition to $\Q$, that is the non-constant embedding tensor, field strength of the vector fields $A_\m{}^{\mc{M}}$ and the anholonomy coefficients for the metric in the external space-time. Bianchi identities for the latter have the same form as in the standard general relativity, while identities for $\Q$ and the field strengths have been known in the literature. Since Bianchi identities explicitly define potentials interacting with magnetically charged branes (at least at the linear level), we use our result to compare to the existing classification of supersymmetric branes in various dimensions \cite{Bergshoeff:2010xc,Bergshoeff:2011zk,Bergshoeff:2011mh,Bergshoeff:2011ee,Kleinschmidt:2011vu}. In addition to states presented in \cite{Kleinschmidt:2011vu} we find exotic branes with special directions along the external space-time. 

Bianchi identities, being identically satisfied by definition, can be understood as a set of condition for a field strength to be defined in terms of a potential. In the simplest examples of Maxwell theory the identity $dF=0$ allows to introduce a gauge potential by $F=dA$. More involved is the example of general relativity, where the Einstein-Hilbert action can be written in terms of anholonomy coefficients, that must satisfy
\begin{equation}
    \dt_{[\m}f_{\n\r]}{}^\s - f_{[\m\n}{}^\k f_{\r]\k}{}^{\s}=0.
\end{equation}
This can be solved in terms of the vielbein $e_{\m}{}^a$ as
\begin{equation}
    f_{\m\n}{}^\r = 2 e_a{}^\r\dt_{[\m}e_{\n}{}^a.
\end{equation}
The metric is then $g_{\m\n}=e_{\m}{}^ae_{\n}{}^b \h_{ab}$. Now, if one considers a transformation of the coefficients $f_{\m\n}{}^\r \to f_{\m\n}{}^\r + \D f_{\m\n}{}^\r$, in order for the new coefficients to be written in terms of (new) vielbeins, they must satisfy Bianchi identities. Precisely this logic applied to generalized anholonomy coefficients (the non-constant embedding tensor $\Q$) has been used in \cite{Bakhmatov:2022rjn,Bakhmatov:2022lin} to arrive at a generalization of 11-dimensional supergravity equations. Supposedly, the Green-Schwarz supermembrane on their solutions is kappa-symmetric. The formalism of \cite{Bakhmatov:2022rjn} is hugely restricted by the condition that all tensor fields of the underlying SL(5) exceptional field theory vanish. Full Bianchi identities derived in this paper serve to extend this formalism to the full SL(5) theory and hence to the full 11D supergravity.

The paper is structured as follows. In Section \ref{sec:BIDFT} we briefly recall Bianchi identities of the full O(10,10) double field theory, where all fields can be packed into two fluxes (and the dilaton). In Section \ref{sec:SL5} we provide all necessary details of exceptional field theory, list weights of all relevant fields, rederive gauge transformations of field strengths and tensor hierarchy fixing typos found in the literature. This section serves as a self-contained description of the SL(5) exceptional field theory. In Section \ref{sec:fluxSL5} we define fluxes of the theory, derive all Bianchi identities for them and provide the full flux Lagrangian. In Section \ref{sec:exotic} we discuss magnetic potential that couple to the listed Bianchi identities. Appendices contain necessary calculational details and the rest can be found in Cadabra files on the GitHub repository \cite{sl5flux:git}.

\section{Fluxes and Bianchi identities in Double field theory}
\label{sec:BIDFT}

In this section we briefly review flux formulation of double field theory and Bianchi identities that are necessary for its consistency. Since double field theory can be formulated in a fully O(10,10)-covariant form, it serves as a convenient setup for illustrating the narrative without going into many technical details. The structure of exceptional field theory with its necessary split into external and internal sets of directions is repeated when the O(10,10) group is broken into $GL(10-d)\times O(d,d)$. This corresponds to $D=10-d$ external directions and $d+d$ doubled internal directions. Here we will be as brief as possible referring for the details to the original papers on double field theory \cite{Siegel:1993xq,Siegel:1993th,Hohm:2010pp,Hohm:2010jy} and its flux formulation \cite{Geissbuhler:2013uka}, and to review papers \cite{Berman:2013eva,Aldazabal:2013sca}. 

In this section we will use the following index conventions:
\begin{equation}
\begin{aligned}
&\a, \b, \g, \dots =1,\dots,6&& \mbox{worldvolume} \\
&i,j,k,l,\dots =1,\dots, d && \mbox{internal space curved} \\
&\ba,\bb,\bc,\bd,\dots =1,\dots, d  && \mbox{internal space flat} \\
&\mu,\nu,\r,\s, \dots =1,\dots, 10-d  && \mbox{external space  curved}\\
&\hat{\mu}, \hat{\nu}, \hat{\r}, \hat{\s},\dots =1,\dots, 10 && \mbox{10D spacetime  curved}\\
& \mc{M}, \mc{N}, \mc{K},\dots = 1,\dots 20 && \mbox{full DFT curved $\rm{O}(10,10)$-covariant}\\
&M,N, K, L, \dots =1,\dots, 2d  && \mbox{DFT curved, ${\rm O}(d,d)$-covariant}\\
&A,B,C,D,\dots =1,\dots, 2d && \mbox{DFT flat, {\rm O}$(d)\times${\rm O}$(d)$-covariant}\\
&a,b,c,d,\dots =1,\dots, d && \mbox{indices labelling Killing vectors}
\end{aligned}
\end{equation}

The full O$(10,10)$-covariant generalized metric of double field theory which reads
\begin{equation}
\hmH_{\mM \mK}=
\begin{bmatrix}
G_{\hmu \hnu} - B_{\hmu \hr} G^{\hrh \hsg} B_{\hsg \hnu} && B_{\hmu \hrh} G^{\hrh \hnu} \\
B_{\hnu \hrh} G^{\hrh \hmu} && G^{\hmu \hnu}.
\end{bmatrix}
\end{equation}
In addition one defines the invariant dilaton $d$ related to the standard dilaton $\f$ by
\begin{equation}
    d = \f -\fr12 \log \det G_{\hat{\m}\hat{\n}}.
\end{equation}
Start with generalized Lie derivative in O(10,10) theory which on the generalized vielbein defined by $\hat{\mc{H}}_{\mc{M}\mc{N}} =\mE_{\mM}{}^{\mA}\mE_{\mN}{}^{\mB}\hat{\mc{H}}_{\mc{A}\mc{B}} $ takes the following form
\begin{equation}
\mL_V \mE_{\mM}{}^{\mA}=V^\mN \dt_\mN \mE_\mM{}^{\mA}+ \mE_\mN{}^\mA \dt_\mM V^\mN-\mE_\mN{}^\mA \dt^\mN V_\mM.
\end{equation}
 Indices are raised and lowered by the O(10,10) invariant tensor
\begin{equation}
\h_{\mM\mN}=
\begin{bmatrix}
 0 & \mathbf{1}\\
 \mathbf{1} & 0
\end{bmatrix}.
\end{equation}
Generalized Lie derivative acting on the generalized vielbein induces the so-called generalized flux, that is simply anholonomy coefficients for the underlying geometry \cite{Lee:2014mla}. In flat indices the flux can be written as follows
\begin{equation}
    \label{eq:fluxesDFT}
        \begin{aligned}
        \mF^{\mA}{}_{\mB\mC}&=2\mE^{\mA}_M \mE_{[\mB}^\mN\dt_\mN \mE_{\mC]}^\mM+\mE^{\mA}_\mM \h^{\mM\mN}\h_{\mK\mL}\dt_\mN \mE_{[\mB}^\mK \mE_{\mC]}^\mL,\\
        \mF_\mA&=\dt_\mM \mE^\mM_\mA+2\mE^\mM_\mA\dt_\mM d.
\end{aligned}
\end{equation}
This flux identically satisfies the following Bianchi identities:
\begin{equation}
\begin{aligned}
&\mE^M_{[\mA}\dt_\mM\mF_{\mB\mC\mD]}-\fr34\mF^\mE{}_{[\mA\mB}\mF_{|\mE|\mC\mD]} \equiv \mZ_{\mA\mB\mC\mD}=0.
\end{aligned}
\end{equation}
Contracting with the generalized vielbein one can rewrite the flux in curved indices as
\begin{equation}
\mF_{\mM\mN\mK}=3\mE_{\mA[\mM}\dt_\mN\mE_{\mK]}{}^\mA\,,
\end{equation}
which now obeys the following BI 
\be
\label{fullBI}
S_{\mM\mN\mP\mQ} \equiv \dt_{[\mM}\mF_{\mN\mK\mL]}-\fr34\mF_{\mP\mM\mN}\mF_{\mQ\mK\mL}\h^{\mP\mQ}=0\,.
\ee
Components of the generalized flux $\mF_{\mM\mN\mP}$ can be identified with a set of space-time tensors: the three-form $H_{\hat{\mu}\hat{\nu}\hat{\rho}}$, the ``geometric flux'' $\tau_{\hat{\mu}\hat{\nu}}{}^{\hat{\rho}}$, the non-geometric Q-flux, $Q_{\hat{\mu}}{}^{\hat{\nu}\hat{\rho}}$ and the non-geometric R-flux, $R^{\hat{\mu}\hat{\nu}\hat{\rho}}$.

To recover the structure of exceptional field theory one should decompose twenty coordinates $\XX^\mM$ into two sets of $2D$ and $2d$ coordinates denoted by $\XX^{\hM}$ and $\XX^M$ respectively. The former will then trivially decompose into the conventional space-time coordinates $x^\m$ and their duals which will  drop from the picture to fulfill the section constraint. The Bianhci identities will decompose correspondingly resulting in a set of identities with various indices
\begin{equation}
    \begin{aligned}
        & S_{MNKL}, && S_{\m MNK}, && S_{\m\n MN}, && S_{\m\n\r M} && S_{\m\n\r\s}.
    \end{aligned}
\end{equation}
In this case the non-derivative part of $S_{MNKL}$ is nothing but the quadratic constraints of half-maximal $SO(d,d)$ gauged supergravity \cite{Weidner:2006rp}. The same observation for fluxes of exceptional field theory will be used to construct similar Bianchi identities without actual calculation of gauge invariance and explicit construction of flux formulation of the theory.

\section{SL(5) theory}
\label{sec:SL5}

In this Section we turn to the SL(5) exceptional field theory that is a U-duality covariant formulation of supergravity in the split $7+4$. In this case one has 7 external coordinates $x^\m$ and 10 coordinates on the extended space. The field content (to be specified below) now contains external metric, scalars and various tensor fields satisfying the so-called tensor hierarchy. For consistency of further narration we provide a detailed description of symmetries of (the bosonic) SL(5) exceptional field theory mentioning relations between the notations we choose here and those chosen in the literature. In what follows we assume the following index conventions for fields and parameters of the SL(5) theory
\begin{equation}
    \begin{aligned}
       &\hat{\mu}, \hat{\nu},\ldots = 1 \dots 11&& \mbox{eleven directions, curved}; \\
       &\hat{\alpha}, \hat{\beta},\ldots = 1 \dots 11&& \mbox{eleven directions, flat}; \\       
       &\mu, \nu, \rho, \ldots = 1,\dots,7 && \mbox{external  directions (curved) of SL(5) ExFT}; \\
       &\bar{\mu}, \bar{\nu}, \bar{\rho}, \ldots 1,\dots,7 && \mbox{external   directions (flat) of SL(5) ExFT}; \\  
       &k, l, m, n,  \ldots = 1,\dots,4 && \mbox{internal directions (curved)}; \\       
       & a, b, c, d, \ldots = 1,\dots,4 && \mbox{internal directions (flat)}; \\  
       & \mM, \mN, \mK, \mL, \ldots = 1,\dots,10 && \mbox{SL(5) ExFT generalized space (curved)};  \\
       & \mA, \mB, \mC, \mD, \ldots = 1,\dots,10 && \mbox{SL(5) ExFT generalized space (flat)};\\
       & M, N, K, L, \ldots = 1,\dots, 5 && \mbox{fundamental SL(5) (curved)};  \\
       & A, B, C, D, \ldots = 1,\dots, 5 && \mbox{fundamental SL(5) (flat)}.
    \end{aligned}
\end{equation}

Generalized space of SL(5) ExFT is parametrized by coordinates $\mathbb{X}^{\mM}$. In terms of fundamental $\bf{5}$ indices of SL(5) they take form $\mathbb{X}^{MN}=-\mathbb{X}^{NM}$. The transition from $\bf{10}$ to antisymmetric pair of $\bf{5}$ is performed as
\begin{equation}
\begin{aligned}[]
T^{\mM} & \rightarrow T^{MN},\qquad\text{any tensor} , \\
U^{\mM} V_{\mM} & \rightarrow \frac12 U^{MN} V_{MN},\\
\delta^{\mM}_{\mN} & \rightarrow 2 \delta^{MN}_{KL},\qquad \text{only Kronecker}.
\end{aligned}
\end{equation}
The additional 2 multiplier above stands for $\delta^{\mM}_{\mM}=\frac12(2\delta^{MN}_{MN})=\delta^{MN}_{MN}=10$.

Epsilon tensors $\epsilon^{M N P Q R}$ and $\epsilon^{A B C D E}$ are absolutely antisymmetric tensors taking values $\pm1$ and are related as
\begin{equation}\label{relationepsilons}
    \epsilon^{M N P Q R} E^{A}{}_{M} E^{B}{}_{N} E^{C}{}_{P} E^{D}{}_{Q} E^{E}{}_{R} = E \, \epsilon^{A B C D E} = e_{(7)}^{- \frac{5}{14}} \epsilon^{A B C D E}.
\end{equation}

\subsection{Generalized geometry and Lie derivative}

Let us start with properties of field transformations under coordinate transformations, i.e. the generalized Lie derivative. For our conventions for generators of the $sl(5)$ algebra see Appendix \ref{SL5algebra}. Generalized Lie derivative of a vector $V^{\mM}$ of weight $\l[V]$ is given by the following expression 
\begin{multline}\label{GenLieY}
    \mL_{\L}V^{\mM} \equiv [\L,V]^{\mM}_{Dorfman} = \\
    =\L^{\mN}\partial_{\mN}V^{\mM} - V^{\mN}\partial_{\mN}\L^{\mM} + Y^{\mM\mN}\,_{\mL\mK}\partial_{\mN}\L^{\mK} V^{\mL} + \bigg(\lambda[V^{\mM}] - \frac15\bigg) \partial_{\mK}\L^{\mK} V^{\mM} = \\
    = \L^{\mN}\partial_{\mN}V^{\mM} - V^{\mN}\partial_{\mN}\L^{\mM} + \epsilon^{M \mM\mN} \epsilon_{M \mK\mL}\partial_{\mN}\L^{\mK} V^{\mL} + \bigg(\lambda[V^{\mM}] - \frac15\bigg) \partial_{\mK}\L^{\mK} V^{\mM} = \\
    = \L^{\mN}\partial_{\mN}V^{\mM} - 3 \mathbb{P}^{\mM}{}_{\mL}{}^{\mN}{}_{\mK} \partial_{\mN}\L^{\mK} V^{\mL} + \lambda[V^{\mM}] \partial_{\mK}\L^{\mK} V^{\mM},
\end{multline}
which in terms of fundamental \textbf{5} indices ($\mM = [M,N]$) takes the following form for vectors
\begin{equation}
    \begin{aligned}
        \mL_{\L}V^{M N} &= \frac{1}{2} \L^{K L} {\dt}_{K L}{{V}^{M N}}\,  - V^{L N} {\dt}_{L K}{{\L}^{M K}}\, - V^{M L} {\dt}_{L K}{{\L}^{N K}}\,  + \bigg(\frac{2}{5} + \frac{\lambda[V^{\mM}]}{2} \bigg)V^{M N} {\partial}_{K L}{{\L}^{K L}} \\
        & = \frac{1}{2} \L^{K L} {\dt}_{K L}{{V}^{M N}}\,  - V^{L N} {\dt}_{L K}{{\L}^{M K}}\, - V^{M L} {\dt}_{L K}{{\L}^{N K}}\, + \frac{\tilde\lambda[V^{\mM}]}{2} V^{M N} {\partial}_{K L}{{\L}^{K L}}.
    \end{aligned}
\end{equation}
For convenience of calculations we define a reduced weight $\tilde{\lambda}$ as twice of the overall coefficient in from of the term  ${\partial}_{K L}{ {\L}^{K L}}$. For transformations of the generalized vector $V^{MN}$ above we have
\begin{equation}
    \tilde{\l}[V^{\mc{M}}] = \l[V^{\mc{M}}] +\fr45.
\end{equation}
It is the weight $\tilde\l$ that is additive. Note that the numerical shift is not always $4/5$ and depends on index structure of the corresponding tensor. For a field $V^{M}$ transforming in the fundamental of the global SL(5) group we have
\begin{equation}\label{vectGL}
    \mL_{\L}V^{M} = \frac{1}{2} \L^{K L} {\dt}_{K L}{{V}^{M}}\,  - V^{L} {\dt}_{L K}{{\L}^{M K}}\, + \bigg(\frac{1}{5} + \frac{\lambda[V^{\mM}]}{4} \bigg) V^{M} {\partial}_{K L}{{\L}^{K L}},
\end{equation}
and similarly for generalized covectors
\begin{equation}\label{covectGL}
    \mL_{\L} V_{M} = \frac{1}{2} \L^{K L} {\dt}_{K L}{{V}_{M}}\,  + V_{L} {\dt}_{M K}{{\L}^{L K}}\, + \bigg( - \frac{1}{5} + \frac{\lambda[V_{\mM}]}{4} \bigg) V_{M} {\partial}_{K L}{{\L}^{K L}}.
\end{equation}
For tensors of arbitrary rank the expressions are modified accordingly. Finally for a weighted scalar one has
\begin{equation}
\label{Liescalar}
    \mL_{\L}\phi = \frac{1}{2} \L^{K L} {\dt}_{K L}{\phi}\, + \frac{\lambda[\phi]}{2} \phi \, {\partial}_{K L}{{\L}^{K L}}.
\end{equation}
Closure of the generalized Lie derivatives' algebra requires section condition
\begin{equation}\label{2cond}
    Y^{\mM\mN}\,_{\mK\mL} \, \partial_{\mM} \bullet \, \partial_{\mN} \, \bullet = 0, \quad Y^{\mM\mN}\,_{\mK\mL} \, \partial_{\mM} \, \partial_{\mN} \, \bullet = 0,
\end{equation}
where $\bullet$ denotes insertion of any field of the theory.

Since in general derivative $\dt_\m V^M$ is no longer a generalized tensor, one introduces external covariant derivative
\begin{equation}\label{covderext}
    \DD_{\mu} = \partial_{\mu} - \mL_{A_{\mu}},
\end{equation}
which is defined with the help of a generalized connection $A_{\mu}{}^{M N}$, that is simply the gauge field of the corresponding $D=7$ maximal supergravity. As we will discuss further, to define a stress tensor for $A_{\mu}{}^{M N}$ covariant under generalized diffeomorphisms more fields has to be added  to the theory, that appear to be a set of five 2-forms $B_{\mu \nu M}$ and five 3-forms $C_{\mu \nu \rho}{}^{M}$.

The final ingredients are the standard metric  $g_{\mu \nu}$ on the $D=7$ space-time (the external space) and  the generalized metric $\mM^{\mM\mN}$ on the extended space (internal).  The latter can be represented as
\begin{equation}\label{GeneralizedMetrics}
            \mM^{MN, KL} = m^{M K} m^{N L} - m^{M L} m^{N K}, \quad \mM^{M,K} \equiv \mM^{MN,}{}^{K}{}_{N} = 4 m^{M K},
\end{equation}
where $m_{MN} = m_{NM}$ denote the generalized metric in the fundamental representation. For further convenience we will need a generalized metric $M^{MN}$ related to $m^{MN}$ by the following rescaling
\begin{equation}\label{GeneralizedMetrics2}
            m_{MN} = e^{\frac17}_{(7)} M_{MN} = M^{-\frac15} M_{MN}.
\end{equation}
As before metrics defined here can be written in terms of the corresponding vielbeins
\begin{equation}\label{GeneralizedMetrics3}
            m_{M N} = m^{A B} \mE^{A}{}_M \mE^B{}_{N},\quad
            M_{M N} = m^{A B} E^{A}{}_M E^{B}{}_N, \quad g_{\mu \nu} = g_{\bm \bn} \,e^{\bm}{}_{\m} e^{\bn}{}_{\n},
\end{equation}
where the $\mE^{A}{}_M \in SL(5)$ and $E^{A}{}_M \in SL(5) \times \mathbb{R}^{+}$, $E^{A}{}_M = e_{(7)}^{-\frac{1}{14}} \mE^{A}{}_M$, $e_{(7)} = \det \, e^{\bm}{}_{\m}$.

As a result, the SL(5) ExFT bosonic field content reads
\begin{equation}\label{ExFTfields}
    \left\{e^{\bm}{}_{\m},\,E^{A}{}_M,\,A_{\mu}{}^{MN},\,B_{\mu \nu M},\,C_{\mu \nu \rho}{}^{M}\right\}.
\end{equation}
For further reference we list  transformations of the above fields under generalized Lie derivative:
\begin{equation}
\begin{aligned}\label{LieoftensorsExFT}
    \mL_{\Lambda} e^{\bm}{}_{\m} & = \frac12 \L^{K L} \partial_{K L} e^{\bm}{}_{\m} + \frac1{10} e^{\bm}{}_{\m} \partial_{K L} \L^{K L}, \\
    \mL_{\Lambda} E_{C}{}^{M} & = \frac12 \Lambda^{K L} \partial_{K L}{E_{C}{}^{M}} - E_{C}{}^{L} \partial_{L K}{\Lambda^{M K}}
    + \frac14 E_{C}{}^{M} \partial_{K L}{\Lambda^{K L}}, \\
    \mL_{\Lambda} \mE_{C}{}^{M} & = \frac{1}{2} \L^{K L} {\dt}_{K L}{\mE_{C}{}^{M}}\,  - \mE_{C}{}^{L} {\dt}_{L K}{{\L}^{M K}}\, + \frac{1}{5} \mE_{C}{}^{M} {\partial}_{K L}{{\L}^{K L}}, \\
    \mL_{\L} A_{\m}{}^{MN} & = \frac{1}{2} \L^{K L} {\dt}_{K L}{A_{\m}{}^{M N}}\,  - 2 A_{\m}{}^{L [N} {\dt}_{L K}{{\L}^{M] K}}\, + \frac{1}{2} A_{\m}{}^{M N} {\partial}_{K L}{{\L}^{K L}}, \\
    \mL_{\L} B_{\m\n M} & = \frac{1}{2} \L^{K L} {\dt}_{K L}{B_{\m\n M}}\,  + B_{\m\n L} {\dt}_{M K}{{\L}^{L K}}, \\
    \mL_{\L} B_{\m\n}{}^{\mM\mN} & = \L^{\mK} \dt_{\mK} B_{\m\n}{}^{\mM\mN} - 2 B_{\m\n}{}^{\mK(\mN} \dt_{\mK} \L^{\mM)} + 2 Y^{\mQ(\mM}{}_{\mK\mL} B_{\m\n}{}^{\mN)\mL} \dt_{\mQ} \L^{\mK}, \\
    \mL_{\L} C_{\m\n\r}{}^M & = \frac{1}{2} \L^{K L} {\dt}_{K L}{C_{\m\n\r}{}^{M}}\,  - C_{\m\n\r}{}^{L} {\dt}_{L K}{{\L}^{M K}}\, + \frac{1}{2} C_{\m\n\r}{}^{M} {\partial}_{K L}{{\L}^{K L}}, \\
    \mL_{\L} C_{\m\n\r}{}^{\mP,\,\mM\mN} & = \L^{\mK} \dt_{\mK} C_{\m\n\r}{}^{\mP,\,\mM\mN} - 2 C_{\m\n\r}{}^{\mP,\,\mK(\mN} \dt_{\mK} \L^{\mM)} + 2 Y^{\mQ(\mM|}{}_{\mK\mL} C_{\m\n\r}{}^{\mP,\,|\mN)\mL} \dt_{\mQ} \L^{\mK} \\
    & \quad - C_{\m\n\r}{}^{\mK,\,\mM\mN} \dt_{\mK} \L^{\mP} + Y^{\mQ\mP}{}_{\mK\mL} C_{\m\n\r}{}^{\mL,\,\mM\mN} \dt_{\mQ} \L^{\mK}. \\
\end{aligned}
\end{equation}
To make contact with the literature (in particular with \cite{Musaev:2015ces} and earlier works on maximal D=7 supergravities \cite{Samtleben:2005bp}) we have also write transformations for the fields $B_{\m\n}{}^{\mK\mL}$ and $C_{\m\n\r}{}^{\mM,\,\mK\mL}$, that are related to $B_{\mu \nu M}$ and $C_{\m\n\r}{}^M$ as follows
\begin{equation}
\label{proper_fields}
 \begin{aligned}
   B_{\m\n}{}^{\mM\mN} & = B_{\m\n}{}^{NKLP} = \frac{1}{48} B_{\m\n M} \e^{MNKLP}, && B_{\m\n M} = 2 \e_{MNKLP} B_{\m\n}{}^{NKLP}, \\
   C_{\m\n\r}{}^{\mM,\,\mK\mL} & = C_{\m\n\r}{}^{MN,KLRS} = \frac{-1}{288} C_{\m\n\r}{}^{[M} \epsilon^{N]KLRS}, && C_{\m\n\r}{}^M = - 6 \e_{NKLRS} C_{\m\n\r}{}^{MN,KLRS}.
 \end{aligned}
\end{equation}

It is important to note, that the generalized Lie derivatives above are taken along generalized vectors $\L,\L'$, which themselves transform as
\begin{equation}
    \mL_{\L'} \L^{M N} = \frac{1}{2} \L'^{K L} {\dt}_{K L}{\L^{M N}}\,  - 2 \L^{L [N} {\dt}_{L K}{{\L'}^{M] K}}\, + \frac{1}{2} \L^{M N} {\partial}_{K L}{{\L'}^{K L}}.
\end{equation}
In the following section generalized vectors with the same transformation properties will be used to define generalized diffeomorphisms and eventually generalized fluxes. In addition, to define external diffeomorphisms we will need scalars $\xi^{\m}$, for which
\begin{equation}
    \mL_{\L} \xi^{\m} = \frac12 \L^{KL}\dt_{KL} \xi^{\m}.
\end{equation}

Finally, we list weights of all fields and parameters of the theory w.r.t. generalized Lie derivatives in Table \ref{tab:weights}.
\begin{table}[http]
    \centering
    \begin{tabular}{|c|c|c|c|c|c|c|c|c|c|c|c|c|}
    \hline
           & $e^{\bm}{}_{\m}$  & $E_{AB}{}^{MN}$ & $\mE_{AB}{}^{MN}$ & $A_{\mu}{}^{MN}$ & $B_{\mu \nu M}$ & $B_{\m\n}{}^{\mK\mL}$ & $C_{\m\n\r}{}^M$ & $C_{\m\n\r}{}^{\mM,\,\mK\mL}$ & $\L^{MN}$ & $\xi^{\mu}$\\
         \hline
         $\lambda$ & $\frac{1}{5}$ & $\frac{1}{5}$ & $0$ & $\frac{1}{5}$ & $\frac45$ & $\frac25$ & $\frac65$ & $\frac35$ & $\frac{1}{5}$ & $0$ \\
         \hline
         $\tilde{\lambda}$ & $\frac{1}{5}$ & $1$ & $\frac45$ & $1$ & $0$ & $2$ & $1$ & $3$ & $1$ & $0$ \\
         \hline
    \end{tabular}
    \caption{Weights under generalized Lie derivative. Here generalized vielbeins are $E_{AB}{}^{MN} = E_{A}{}^{[M} E_{B}{}^{N]} \in SL(5)\times\mathbb{R}^{+}$ and $\mE_{AB}{}^{MN} = \mE_{A}{}^{[M} \mE_{B}{}^{N]} \in SL(5)$, $\L^{MN}$ is generalized vector used to build generalized Lie derivatives and diffeomorphisms, and $\xi^{\m}$ is scalar used to build external diffeomorphisms. $\l$ is the weight used in formulas (\ref{GenLieY})-(\ref{Liescalar}), and $\tilde{\lambda}$ is the reduced weight - the number standing in front of the term $\dt_{\mK} \L^{\mK}$ in generalized lie derivative, after all simplifications (as in (\ref{LieoftensorsExFT})). It is the weight $\tilde{\lambda}$ that is additive.}
    \label{tab:weights}
\end{table}

\subsection{Generalized diffeomorphisms and tensor hierarchy}

Similarly to what happens in gauged supergravities commuting external covariant derivatives $\mc{D}_\m$ does not lead to a covariant field strength
\begin{equation}
    [\mD_{\m},\mD_{\n}] = - \mL_{F_{\m\n}}.
\end{equation}
To covariantize the derived expression $F_{\m\n}$ one uses the 2-form gauge potential entangling its transformations with those of the 1-form. The idea is to add to $F_{\m\n}$ a term proportional to the so-called trivial generalized Lie parameter, i.e. such that does not change the RHS above.  Next, to construct a field strength 2-form that is a tensor under generalized Lie derivatives one uses 3-form and so on. The corresponding hierarchy of tensor fields has been observed to be inevitable in \cite{deWit:2005hv,deWit:2008ta} for gauged supergravities and in \cite{Hohm:2013pua} for exceptional field theories. Let us list the resulting expressions for the covariant field strengths\footnote{Note that in reference \cite{Musaev:2015ces} there is a typo in formula (3.13) in expression for $\mF_{\m\n}{}^{MN}$: there should be  a ``-'' sign in front of the term with B-field.}
\begin{equation}
\label{hierarchy}
\begin{aligned}
\mF_{\m\n}{}^{MN} =&\ 2 \dt_{[\m}A_{\n]}{}^{MN} - [A_{\m},A_{\n}]_E^{MN} - \frac1{16} \epsilon^{M N K L P} \dt_{K L} B_{\m \n P},\\
\mF_{\m \n \r M}  =&\ 3 \, \DD_{[\m} B_{\n \r] M} + 6 \, \epsilon_{M P Q R S} (A_{[\mu}{}^{P Q} \dt_{\nu} A_{\rho]}{}^{R S} - \frac13 [A_{[\mu},A_{\nu}]_{E}^{P Q} A_{\rho]}{}^{R S}) - \dt_{M N} C_{\m \n \r}{}^{N}, \\
\mF_{\m\n\r}{}^{\mK\mL}  = &\ 3\,\mc{D}_{[\m}B_{\n\r]}{}^{\mK\mL} +
 \fr{1}{2} \, Y^{\mK\mL}{}_{\mP\mQ}\Big(A_{[\m}{}^{\mP}\dt_\n
A_{\r]}{}^{\mQ}-\fr13 [A_{[\m},A_{\n}]_E{}^{\mP}A_{\r]}{}^{\mQ}\Big)\\
& - 3 \big( \dt_\mN C_{\m\n\r}{}^{\mN,\mK\mL} - Y^{\mK\mL}{}_{\mP\mQ} \, \dt_\mN C_{\m\n\r}{}^{\mQ,\mP\mN} \big), \\
\mF_{\m\n\r\s}{}^{M} =&\ 
4\,\mc{D}_{[\m}C_{\n\r\s]}{}^{M} - 6 \mF_{[\m\n}{}^{MN} B_{\r\s] N} - \frac{3}{16} \e^{MNKLP} B_{[\m\n| N} \dt_{KL} B_{|\r\s] P} \\
& - 32 \e_{N \mP\mQ} \left(A_{[\m}{}^{MN} A_\n{}^\mP \dt_\r A_{\s]}{}^\mQ - \fr14 A_{[\m}{}^{MN} [A_\n, A_\r]_E{}^\mP 
A_{\s]}{}^\mQ\right) \\
& + \frac12 \e^{M K L P Q} \dt_{K L} \mG_{\m\n\r\s P Q},\\
\mF_{\m\n\r\s}{}^{MN,\mK\mL} =&\ 
4 \, \mD_{[\m} C_{\n\r\s]}{}^{MN,\mK\mL} + \left(2 B_{[\m\n}{}^{\mK\mL} \mF_{\r\s]}{}^{MN} +  
Y^{MN\mN}{}_{\mP\mQ} B_{[\m\n}{}^{\mK\mL} \dt_\mN B_{\r\s]}{}^{\mP\mQ}\right)\\
& + \fr{2}{9} 
Y^{\mK\mL}{}_{\mP\mQ} \left(A_{[\m}{}^{MN} A_\n{}^\mP \dt_\r A_{\s]}{}^\mQ - \fr14 A_{[\m}{}^{MN} [A_\n, A_\r]_E{}^\mP 
A_{\s]}{}^\mQ\right) \\
& - \frac{1}{144} \e^{\mK \mL [N} \e^{M] \mP \mQ} \dt_{\mP} \mG_{\m\n\r\s \mQ},\\
\mF_{\mu\nu\rho\sigma\tau M N} =&\ \ \, 5\,\mD_{[\mu} \mG_{\nu\rho\sigma\tau] MN} + \frac{15}{16}\, B_{[\mu\nu| M} \mD_\rho B_{\sigma\tau] N} - \frac{15}{16}\, B_{[\mu\nu| M} \mD_\rho B_{\sigma\tau] N} \\
&- \frac{5}{2}\, \e_{MNKLP} \mc{F}_{[\mu\nu}{}^{K L} C_{\rho\sigma\tau]}{}^{P} + ...
\end{aligned}
\end{equation}
where the E-bracket is given by
\begin{equation}
    [\L_{1},\L_{2}]_{E} = \frac12 [\L_{1},\L_{2}]_{Dorfman} - \frac12 [\L_{2},\L_{1}]_{Dorfman}.
\end{equation}
The fields $\mF_{\m\n\r}{}^{\mK\mL}$, $\mF_{\m\n\r\l}{}^{\mM,\,\mK\mL}$ and $\mF_{\m\n\r M}$, $\mF_{\m\n\r\l}{}^{M}$ are in the same relations as (\ref{proper_fields}). The 4-form $\mG_{\m\n\r\s P Q}$ is an auxiliary field introduced to make the field strength for $C_{\m\n\r}{}^{M}$ covariant under generalized diffeomorphisms. Its  stress tensor $\mF_{\mu\nu\rho\sigma\tau M N}$ does not contribute to the SL(5) ExFT action and is needed only for Bianchi identities for $\mF_{\m\n\r\s}{}^{M}$
\begin{equation}
    5 \, \DD_{[\m} \mF_{\n \r \s \tau]}{}^{M} =  \frac{1}{2} \, \epsilon^{M N P K L} \dt_{N P} \mF_{\m \n \r \s \tau K L} - 10 \mF_{[\m \n}{}^{M N} \mF_{\r \sigma \tau] N}.
\end{equation}
The complete expression for $\mF_{\mu\nu\rho\sigma\tau M N}$ can be found in \cite{Wang:2015hca}. 

Under arbitrary variations $\d A_\m{}^{MN}$, $\d B_{\m\n M}$, $\d B_{\m\n}{}^{\mK\mL}$, $\d C_{\m\n\r}{}^{M}$, $\d C_{\m\n\r}{}^{\mN,\mK\mL}$, $\d G_{\mu \nu \rho \sigma M N}$ of the $p$-form potentials the covariant field strengths transform as 
follows
\begin{equation}
\label{varF}
\begin{aligned}
\d \mF_{\m\n}{}^{MN}= & 2\,\mc{D}_{[\m}\D A_{\n]}{}^{MN} - \frac1{16} \epsilon^{M N K L P} \dt_{K L} \D B_{\m\n P},\\
\d \mF_{\m \n \r M}  =&\ 3 \, \DD_{[\m} \D B_{\n \r] M} + 6 \, \epsilon_{M P Q R S} \mF_{[\m \n}{}^{P Q} \d A_{\r]}{}^{R S} - \dt_{M N} \D C_{\m \n \r}{}^{N}, \\
\d \mF_{\m\n\r}{}^{\mK\mL}=&\ 3\,\mc{D}_{[\m}\D 
B_{\n\r]}{}^{\mK\mL} + \fr{1}{2} Y^{\mK\mL}{}_{\mP\mQ}\,\mF_{[\m\n}{}^\mP \D 
A_{\r]}{}^\mQ \\
& - 3 \big( \dt_\mN \D C_{\m\n\r}{}^{\mN,\mK\mL} - Y^{\mK\mL}{}_{\mP\mQ}\, \dt_\mN \D C_{\m\n\r}{}^{\mQ,\mP\mN} \big),\\
\d \mF_{\m\n\r\s}{}^{M} =&\ 4\, D_{[\m}\D C_{\n\r\s]}{}^{M} - 6 \mF_{[\m\n}{}^{MN} \D B_{\r\s] N} + 4 \mF_{[\m\n\r| N} \d A_{|\s]}{}^{MN} \\
 &+ \frac12 \e^{M K L P Q} \dt_{K L} \D \mG_{\m\n\r\s P Q}, \\
\d \mF_{\m\n\r\s}{}^{\mM,\mK\mL} =&\ 4\, D_{[\m}\D C_{\n\r\s]}{}^{\mM,\mK\mL} + 2 \mF_{[\m\n}{}^{\mM}\D B_{\r\s]}{}^{\mK\mL} - \fr{4}{3} \mF_{[\m\n\r}{}^{\mK\mL} \d A_{\s ]}{}^\mM \\
&- \frac{1}{144} \e^{\mK \mL [N} \e^{M] \mP \mQ} \dt_{\mP} \D \mG_{\m\n\r\s \mQ}.
\end{aligned}
\end{equation}
where it proves useful to introduce the following ``covariant'' transformations
\footnote{When comparing the calculations to  the same expressions in the literature we have found numerous typos, which we neatly fix here. First in \cite{Musaev:2015ces} the sign for the C-field transformations in (3.16), (3.19) there must be the opposite. Note that the factors in the last line of (3.15) are different. Also we find typos in (4.58), (4.62) of \cite{Berman:2020tqn} in indices of terms of the form $\delta_{\mM}^{\mN} Y^{\mK\mL}{}_{\mP\mQ}$.

In \cite{Berman:2020tqn,Wang:2015hca} the expression for $\D \mG_{\m\n\r\s P Q}$ should have the opposite signs in front of $B \d B$ term and in front of $B \d A$. To compare our expressions to those of \cite{Berman:2020tqn,Wang:2015hca} one has to perform field redefinitions $B_{\m\n}\rightarrow -8B_{\m\n}$, $C_{\m\n\r}\rightarrow -8C_{\m\n\r}$, $\mG_{\m\n\r\s}\rightarrow -8D_{\m\n\r\s}$.}$^,$
\footnote{Explicit calculations leading to (\ref{varF}) - (\ref{strensthtensortransform}) are collected in Cadabra files \texttt{FAB\_variation}, \texttt{Lie\_F4}, \texttt{Var\_F4(1)} and \texttt{Var\_F4(2)}  of \cite{sl5flux:git}}
\begin{equation}
\begin{aligned}\label{covariantvariationsfields}
\D A_\m{}^{MN} =&\ \d A_\m{}^{MN},\\
\D B_{\m\n M} =&\ \d B_{\m\n M} - 2 \e_{MNKLP}A_{[\m}{}^{NK}\d A_{\n]}{}^{LP},\\
\D B_{\m\n}{}^{\mK\mL} =&\ \d B_{\m\n}{}^{\mK\mL} - \fr{1}{6} Y^{\mK\mL}{}_{\mM\mN} A_{[\m}{}^{\mM} \d A_{\n]}{}^{\mN},\\
\D C_{\m\n\r}{}^{M} =&\  \d C_{\m\n\r}{}^{M} - 3\d A_{[\m}{}^{MN} B_{\n\r] N} - 2 \e_{NKLRS} 
A_{[\m}{}^{MN} A_\n{}^{KL} \d A_{\r]}{}^{RS}, \\
\D C_{\m\n\r}{}^{\mN,\mK\mL} =&\  \d C_{\m\n\r}{}^{\mN,\mK\mL} + \d A_{[\m}{}^\mN B_{\n\r]}{}^{\mK\mL} + 
\fr{1}{18} Y^{\mK\mL}{}_{\mR\mS} A_{[\m}{}^\mN A_\n{}^\mR \d A_{\r]}{}^\mS, \\
\D \mG_{\m\n\r\s M N} =&\ \d G_{\mu \nu \rho \sigma M N} - \epsilon_{M N Q P K} \d A_{\mu}{}^{Q P} C_{\nu \rho \sigma}{}^{K} + \frac{3}{16} B_{[\mu \nu| M} \d B_{|\rho \sigma] N} - \frac{3}{16} B_{[\mu \nu| N} \d B_{|\rho \sigma] M} \\
& - \frac{3}{2} \epsilon_{N K L P R} B_{[\mu \nu| M} A_{|\rho}{}^{K L} \d A_{\sigma]}{}^{P R} - \frac{1}{2} \epsilon_{M N U V R} \epsilon_{Q K L S T} A_{[\mu}{}^{U V} A_{\nu}{}^{R Q} A_{\rho]}{}^{K L} \d A_{\sigma}{}^{S T}.
\end{aligned}
\end{equation}
The field strengths defined above are covariant under generalized diffeomorphisms (generalized Lie derivative)
\begin{equation}
\begin{aligned}
    \delta_{\Lambda} E_{C}{}^{M} & = \mL_{\Lambda} E_{C}{}^{M} = \frac12 \Lambda^{K L} \partial_{K L}{E_{C}{}^{M}} - E_{C}{}^{L} \partial_{L K}{\Lambda^{M K}}
    + \frac14 E_{C}{}^{M} \partial_{K L}{\Lambda^{K L}}, \\
    \delta_{\Lambda} e^{\bm}{}_{\m} & = \mL_{\Lambda} e^{\bm}{}_{\m} = \frac12 \L^{K L} \partial_{K L} e^{\bm}{}_{\m} + \frac1{10} e^{\bm}{}_{\m} \partial_{K L} \L^{K L}, \\
    \delta_{\L} A_{\m}{}^{MN} & = \DD_{\m} \L^{MN} = \dt_{\mu} \L^{MN} - \mL_{A_{\mu}} \L^{MN}, \\
    \delta_{\L} B_{\m\n M} & = - 2 \e_{MNKLP}\L^{NK} \mF_{\m\n}{}^{LP},\\
    \delta_{\L} B_{\m\n}{}^{\mK\mL} & = - \fr{1}{6} Y^{\mK\mL}{}_{\mM\mN} \L^\mM\mc{F}_{\m\n}{}^\mN, \\
    \delta_{\L} C_{\m\n\r}{}^M & = \L^{M N} \mF_{\m\n\r N}, \\
    \delta_{\L} C_{\m\n\r}{}^{\mM,\mK\mL} & = \fr{1}{9} Y^{\mK\mL}{}_{\mP\mQ} \L^\mP \mF_{\m\n\r}{}^{\mQ\mM}, \\
    \d_{\L} \mG_{\m\n\r\s M N} & = \frac{1}{4} \e_{K L M N P} \L^{M N} \mF_{\m\n\r\s}{}^{P},
    \end{aligned}
\end{equation}
and transform as tensors\footnote{Interesting to mention that to prove (\ref{strensthtensortransform}) one needs  Bianchi identities that will appear further in (\ref{BIfirst}), more specifically one uses the first three of them. This implies, that in presence of objects, magnetically charged with respect to the fields of (\ref{strensthtensortransform}), invariance under generalized diffeomorphisms will be broken by terms proportional to the corresponding sources.}
\begin{equation}
\label{strensthtensortransform}
\begin{aligned}
    \d_{\L} \mF_{\m\n}{}^{MN} & = \mL_{\L} \mF_{\m\n}{}^{MN} = \frac{1}{2} \L^{K L} {\dt}_{K L}{\mF_{\m\n}{}^{M N}}\,  - 2 \mF_{\m\n}{}^{L [N} {\dt}_{L K}{{\L}^{M] K}}\, + \frac{1}{2} \mF_{\m\n}{}^{M N} {\partial}_{K L}{{\L}^{K L}}, \\
    \d_{\L} \mF_{\m\n\r M} & = \mL_{\L} \mF_{\m\n\r M} = \frac{1}{2} \L^{K L} {\dt}_{K L}{\mF_{\m\n\r M}}\,  + \mF_{\m\n\r L} {\dt}_{M K}{{\L}^{L K}}, \\
    \d_{\L} \mF_{\m\n\r}{}^{\mM\mN} & = \mL_{\L} \mF_{\m\n\r}{}^{\mM\mN} = \L^{\mK} \dt_{\mK} \mF_{\m\n\r}{}^{\mM\mN} - 2 \mF_{\m\n\r}{}^{\mK(\mN} \dt_{\mK} \L^{\mM)} + 2 Y^{\mQ(\mM}{}_{\mK\mL} \mF_{\m\n\r}{}^{\mN)\mL} \dt_{\mQ} \L^{\mK}, \\
    \d_{\L} \mF_{\m\n\r\l}{}^M & = \mL_{\L} \mF_{\m\n\r\l}{}^M = \frac{1}{2} \L^{K L} {\dt}_{K L}{\mF_{\m\n\r\l}{}^{M}}\,  - \mF_{\m\n\r\l}{}^{L} {\dt}_{L K}{{\L}^{M K}}\, + \frac{1}{2} \mF_{\m\n\r\l}{}^{M} {\partial}_{K L}{{\L}^{K L}}, \\
    \d_{\L} \mF_{\m\n\r\l}{}^{\mP,\,\mM\mN} & = \mL_{\L} \mF_{\m\n\r\l}{}^{\mP,\,\mM\mN} = \L^{\mK} \dt_{\mK} \mF_{\m\n\r\l}{}^{\mP,\,\mM\mN} - 2 \mF_{\m\n\r\l}{}^{\mP,\,\mK(\mN} \dt_{\mK} \L^{\mM)} - \mF_{\m\n\r\l}{}^{\mK,\,\mM\mN} \dt_{\mK} \L^{\mP} \\
    & \quad + 2 Y^{\mQ(\mM|}{}_{\mK\mL} \mF_{\m\n\r\l}{}^{\mP,\,|\mN)\mL} \dt_{\mQ} \L^{\mK} + Y^{\mQ\mP}{}_{\mK\mL} \mF_{\m\n\r\l}{}^{\mL,\,\mM\mN} \dt_{\mQ} \L^{\mK}, \\
\end{aligned}
\end{equation}
that can be summarized as
\begin{table}[http]
    \centering
    \begin{tabular}{|c|c|c|c|c|c|c|c|}
    \hline
           & $\mF_{\mu \n}{}^{MN}$ & $\mF_{\mu \nu \r M}$ & $\mF_{\m\n\r}{}^{\mK\mL}$ & $\mF_{\m\n\r\l}{}^M$ & $\mF_{\m\n\r\l}{}^{\mM,\,\mK\mL}$  \\
         \hline
         $\lambda$ & $\frac{1}{5}$ & $\frac45$ & $\frac25$ & $\frac65$ & $\frac35$ \\
         \hline
         $\tilde{\lambda}$ & $1$ & $0$ & $2$ & $1$ & $3$  \\
         \hline
    \end{tabular}
    \caption{Weights under generalized Lie derivative. $\l$ is the weight used in formulas (\ref{GenLieY})-(\ref{Liescalar}), and $\tilde{\lambda}$ is the reduced weight.}
    \label{tab:weightsF}
\end{table}

\subsection{Gauge transformations}

Gauge transformations parametrised by $\X_{\m M}$ and $\Y_{\m\n}{}^{N}$ are constructed such that the defined above 2-, 3- and 4-form field 
strengths transform covariantly. Explicitly we have 
\begin{equation}
\begin{aligned}
\D_{\X_{\m}, \Y_{\m\n}} A_\m{}^{MN} =&\ \frac{1}{16} \epsilon^{MNKLP}\dt_{KL}\X_{\m P},\\
\D_{\X_{\m}, \Y_{\m\n}} B_{\m\n M} =&\ 2 \mc{D}_{[\m}\X_{\n] M} - \dt_{N M} \Y_{\m\n}{}^{N},\\
\D_{\X_{\m}, \Y_{\m\n}} B_{\m\n}{}^{\mK\mL} =&\ 2\mc{D}_{[\m}\X_{\n]}{}^{\mK\mL} + 3 \left(\dt_\mN\Y_{\m\n}{}^{\mN,\mK\mL} - Y^{\mK\mL}{}_{\mP\mQ} \dt_\mN \Y_{\m\n}{}^{\mP,\mN\mQ}\right),\\
\D_{\X_{\m}, \Y_{\m\n}} C_{\m\n\r}{}^{M} =&\ 3 \mc{D}_{[\m} \Y_{\n\r]}{}^{M} + 3 \mc{F}_{[\m\n}{}^{MN} \X_{\r]N}, \\
\D_{\X_{\m}, \Y_{\m\n}} C_{\m\n\r}{}^{\mM,\mK\mL} =&\ 3 \mc{D}_{[\m} \Y_{\n\r]}{}^{\mM,\mK\mL} - \mc{F}_{[\m\n}{}^{\mM} \X_{\r]}{}^{\mK\mL}. \\
 \end{aligned}
\end{equation}
Relations between different representations of gauge parameters read
\begin{equation}
\label{proper_gaugerel}
 \begin{aligned}
   \X_{\m}{}^{\mM\mN} & = \X_{\m}{}^{NKLP} = \frac{1}{48} \X_{\m M} \e^{MNKLP}, && \X_{\m M} = 2 \e_{MNKLP} \X_{\m}{}^{NKLP}, \\
   \Y_{\m\n}{}^{\mM,\,\mK\mL} & = \Y_{\m\n}{}^{MN,KLRS} = \frac{-1}{288} \Y_{\m\n}{}^{[M} \epsilon^{N]KLRS}, && \Y_{\m\n}{}^M = - 6 \e_{NKLRS} \Y_{\m\n}{}^{MN,KLRS}.
 \end{aligned}
\end{equation}

\subsection{External diffeomorphisms}

Finally, local transformations of the bosonic SL(5) exceptional field theory include external diffeomorphisms $x^{\m} \rightarrow x^{\m} + \x^{\m}(x^{\m}, \mathbb{X}^{MN})$
\begin{equation}
 \begin{aligned}
  \d_{\xi} e^{\bm}{}_\m &= \x^\n \DD_\n e^{\bm}{}_\m + e^{\bm}{}_\n \DD_\m \x^\n ,\\
  \d_{\xi} \mE^{A}{}_{M} &= \x^\m \mc{D}_\m \mE^{A}{}_{M},\\
    \d_{\xi} E^{A}{}_{M} &= \x^\m \mc{D}_\m E^{A}{}_{M} - \frac1{14} E^{A}{}_{M} \DD_{\m} \x^{\m},\\
   \d_{\xi} A_\m{}^{MN} &= \x^\n \mF_{\n\m}{}^{MN} + \frac12 \M^{MN,KL} g_{\m\n} \dt_{KL} \x^\n,\\
     \D_{\xi} B_{\m\n M} &= \x^\r \mF_{\r\m\n M},\\
   \D_{\xi} C_{\m\n\r}{}^{M} &= -\fr{1}{3!} \ve_{\m\n\r\s\k\l\t} \x^\s m^{M N} \mF^{\k\l\t}{}_{N} = \xi^{\sigma} \mF_{\s\m\n\r}{}^{M},
 \end{aligned}
\end{equation}
where $\ve_{\m\n\r\s\k\l\t}=e_{(7)} \e_{\m\n\r\s\k\l\t}$ is the Levi--Civita tensor in 7 dimensions. Note that the last equality here follows from the duality (\ref{BCduality}) assuming that the external metric $g_{\m\n}$ has negative signature $\mathrm{sign}[g_{\m\n}] = - 1$. Apparently, for the internal space this implies $\mathrm{sign}[g_{mn}] = + 1$.

\section{Flux formulation of SL(5) ExFT}
\label{sec:fluxSL5}

As field equations of general relativity can be written in terms of anholonomy coefficients, equations of double field theory can be written in terms of generalized fluxes \cite{Geissbuhler:2013uka}. This is what is known under the name of flux formulation and appears to a convenient representation of 10d supergravity equations for solution generation techniques based on T-dualities \cite{Borsato:2020bqo}, or for constructing equations of the 10d Type II generalized supergravity.

Flux formulation of double field theory of \cite{Geissbuhler:2013uka} cannot be immediately rewritten for exceptional field theories due to the necessary split between the external (ordinary space-time) and internal (extended) directions. Certain similarity can be achieved in the truncation when all tensor fields vanish and dependence of the external metric on the extended coordinates factorizes \cite{Blair:2014zba}. This has been used in \cite{Gubarev:2020ydf} to construct polyvector deformations and in \cite{Bakhmatov:2022lin} to construct a generalization of  supergravity in 11 dimensions.

We now proceed with a construction of flux formulation  of the SL(5) exceptional field theory, where by fluxes we will mean i) anholonomy coefficients for the external metric, ii) field strengths of gauge fields, iii) scalar generalized fluxes without external indices, iv) few other necessary objects encoding fluxes sourced by non-geometric branes spanning partially the ordinary space-time.

\subsection{Fluxes of SL(5) ExFT}

We define the full set of SL(5) ExFT fluxes enough to encode its field equations as follows\footnote{Note, that in Cadabra calculations we sometimes use $\mF_{M N} = e_{(7)}^{-1} \dt_{M N} e_{(7)}$, which is neither a flux nor an SL(5) tensor, and has been introduced merely for convenience.}
\begin{equation}\label{sl5fluxes}
\begin{aligned}
\mF_{\m \n}{}^{M N} & = 2 \dt_{[\m}A_{\n]}{}^{MN} - [A_{\m},A_{\n}]_E^{MN} - \frac1{16} \epsilon^{M N K L P} \dt_{K L} B_{\m \n P}, \\
\mF_{\m \n \r M} & = 3 \, \DD_{[\m} B_{\n \r] M} + 6 \, \epsilon_{M P Q R S} (A_{[\mu}{}^{P Q} \dt_{\nu} A_{\rho]}{}^{R S} - \frac13 [A_{[\mu},A_{\nu}]_{E}^{P Q} A_{\rho]}{}^{R S}) - \dt_{M N} C_{\m \n \r}{}^{N}, \\
\mF_{\m\n\r\s}{}^{M} & = 
4\,\mc{D}_{[\m}C_{\n\r\s]}{}^{M} - 6 \mF_{[\m\n}{}^{MN} B_{\r\s] N} - \frac{3}{16} \e^{MNKLP} B_{[\m\n| N} \dt_{KL} B_{|\r\s] P} \\
& - 32 \e_{N \mP\mQ} \left(A_{[\m}{}^{MN} A_\n{}^\mP \dt_\r A_{\s]}{}^\mQ - \fr14 A_{[\m}{}^{MN} [A_\n, A_\r]_E{}^\mP 
A_{\s]}{}^\mQ\right) + \frac12 \e^{M K L P Q} \dt_{K L} \mG_{\m\n\r\s P Q},\\
\mF_{\bm \bn}{}^{\bl} & = 2 \, e_{[\bm}{}^{\m} e_{\bn]}{}^{\n} \DD_{\m} e^{\bl}{}_{\n},\\
{\mF^{(E)}}_{\m A}{}^{B} & = E_{A}{}^{M} \DD_{\m} E^{B}{}_{M} = \mF_{\m A}{}^{B} - \frac{1}{14} \delta_{A}{}^{B}  e_{(7)}^{-1} \DD_{\m} e_{(7)} = \mF_{\m A}{}^{B} - \frac{1}{14} \delta_{A}{}^{B} G_{\mu}, \\
{\mF}_{M N \bm}{}^{\bn} & = e_{\bm}{}^{\m} \dt_{M N} e^{\bn}{}_{\m} - \frac17 \,  \delta_{\bm}{}^{\bn} e_{(7)}^{-1} \dt_{M N} e_{(7)} = e_{\bm}{}^{\m} e_{(7)}^{\frac17} \dt_{M N}(e_{(7)}^{-\frac17} e^{\bn}{}_{\m}), \\
\mF_{ABC}{}^{D} & =  \frac32 E^{D}{}_{N} \partial_{[A B} E_{C]}{}^{N} - E_{C}{}^{M} \partial_{M N} E_{[B}{}^{N} \delta^{D}{}_{A]} -  \frac12 E_{[B|}{}^{M} \partial_{M N} E_{|A]}{}^{N} \delta^{D}{}_{C}.
\end{aligned}
\end{equation}
here we use $E^{A}{}_M \in SL(5) \times \mathbb{R}^{+}$, $E^{A}{}_M = e_{(7)}^{-\frac{1}{14}} \mE^{A}{}_M$, $e_{(7)} = \det \, e^{\bm}{}_{\m}$. For example, some of the above fluxes enter the definition of the generalized (improved) Riemann tensor
\begin{equation}\label{genRieman}
    \begin{aligned}
    \hat{R}_{\mu\nu}{}^{\bm\bn} & =  R_{\mu\nu}{}^{\bm\bn} + \mF_{\mu\nu}{}^{M N} e^{\bm}{}^{\rho}\partial_{M N} e^{\bn}{}_{\rho}\, = R_{\mu\nu}{}^{\bm\bn} + \mF_{\mu\nu}{}^{M N} {\mF_{M N}{}^{[\bm \bn]}}\, \\
    R_{\m\nu}{}^{\bm\bn} & = 2 \mD_{[\m} \omega_{\n]}{}^{\bm\bn} + 2 \omega_{[\m}{}^{\bm\br} \, \omega_{\n] \br}{}^{\bn}, \quad 
    \omega_{\m}{}^{\bm\bn} = \frac12 e_{\br \m} (\mF^{\bm \bn \br} - \mF^{\bn \br \bm} - \mF^{\br \bm \bn}),
    \end{aligned}
\end{equation}
 and Ricci scalar 
\begin{equation}\label{genRicci}
    \hat{R} = \hat{R}_{\mu\nu}{}^{\m\n}= R + \mF_{\mu\nu}{}^{M N} \mF_{M N}{}^{\m \n},
\end{equation}
where the flat indices are lowered and raised with flat external metric $g_{\bm \bn}$, $g^{\bm \bn}$.  Note that $\mF_{M N}{}^{\m \n}$ is \emph{not related} to the field strength $\mF_{\m \n}{}^{M N}{}$  by raising and lowering of indices. It proves convenient for calculations to introduce additional combinations of fluxes
\begin{equation}\label{sl5fluxesauxil}
\begin{aligned}
\mF_{\bm} &  = \DD_{\m} e_{\bm}{}^{\m} = \mF_{\bm \n}{}^{\n} + \frac{14}{5} {\mF^{(E)}}_{\bm A}{}^{A},\\
G_{\m} &  = e_{(7)}^{-1} \DD_{\m} e_{(7)} = \mF_{\m \n}{}^{\n} - \mF_{\m}, \\
\mF_{\m A}{}^{B} & = \mE_{A}{}^{M} \DD_{\m} \mE^{B}{}_{M} = {\mF^{(E)}}_{\m A}{}^{B} + \frac{1}{14} \delta_{A}{}^{B} G_{\mu}.
\end{aligned}
\end{equation}
Explicit check as in \cite{Berman:2012uy} shows that $\mF_{AB,C}{}^D$ contains only components in the $\bf 10$, $\bf 15$ and $\bf \overline{40}$, which are conventionally called $\q_{AB},Y_{AB}$ and $Z^{ABC}$ respectively, i.e. one can write
\begin{equation}\label{internalfluxes}
    \begin{aligned}
        \mF_{ABC}{}^D &= \frac32 Z_{ABC}{}^D + 5 \theta_{[AB}\d_{C]}{}^D +  \d_{[A}{}^{D} Y_{B]C}. 
    \end{aligned}
\end{equation}
Reversely, the irreducible flux components are written as follows 
\begin{equation}
    \begin{aligned}\label{TYZfluxes}
    \q_{AB} &=\frac15 \mF_{A B,C}{}^{C}= \frac{1}{10} E_{[A}{}^M \dt_{MN}E_{B]}{}^N - \frac{1}{10} E^{-1} \, E_{AB}{}^{MN}\dt_{MN}E , \\
    Y_{AB}& =\frac12 \mF_{C(A,B)}{}^{C}= - E_{(A}{}^M \dt_{MN}E_{B)}{}^N ,\\
    Z_{ABC}{}^{D}&= \fr23 \mF_{[ABC]}{}^{D} - 5 \theta_{[AB} \delta_{C]}{}^{D}\\
    &= E_{[A}{}^{M} E_{B|}{}^{N} E^{D}{}_{K} \partial_{M N}{E_{|C]}{}^{K}} + \frac13 \Big(2 E_{[A|}{}^{M} \partial_{M N}{E_{|B|}{}^{N}}  + E_{[A}{}^{M} E_{B|}{}^{N} E^{-1} \partial_{M N}{E}\Big) \delta_{|C]}{}^{D},\\
     Z^{F G E} & = - \frac{1}{16} \e^{FGABC} Z_{ABC}{}^D\\
     &= - \frac{1}{24} \epsilon^{F G A B C} E^{E}{}_{M} \partial_{A B}{E_{C}{}^{M}} - \frac{1}{48} \epsilon^{G A B C E} E^{F}{}_{M} \partial_{A B}{E_{C}{}^{M}} + \frac{1}{48} \epsilon^{F A B C E} E^{G}{}_{M} \partial_{A B}{E_{C}{}^{M}},
    \end{aligned}
\end{equation}
where $E = \det \, E^{A}{}_{M} $.

Finally let us write the transformation of new introduced fluxes under generalized diffeomorphisms  that complement \eqref{strensthtensortransform}
\begin{equation}
\label{strensthtensortransform2}
\begin{aligned}
    \d_{\L} \mF_{\bm \bn}{}^{\bl} & = \mL_{\L} \mF_{\bm \bn}{}^{\bl} = \frac{1}{2} \L^{K L} {\dt}_{K L}{\mF_{\bm \bn}{}^{\bl}}\, - \frac{1}{10} \mF_{\bm \bn}{}^{\bl} {\partial}_{K L}{{\L}^{K L}}, \\
    \d_{\L} {\mF^{(E)}}_{\m A}{}^{B} & = \mL_{\L} {\mF^{(E)}}_{\m A}{}^{B} = \frac{1}{2} \L^{K L} {\dt}_{K L}{{\mF^{(E)}}_{\m A}{}^{B}}\, , \\
    \d_{\L} {\mF}_{M N \bm}{}^{\bn} & = \mL_{\L} {\mF}_{M N \bm}{}^{\bn} = \frac12 \L^{K L} \dt_{K L} {\mF}_{M N \bm}{}^{\bn} + \frac12 \mF_{K L \bm}{}^{\bn} \partial_{M N}{\Lambda^{K L}} \\
    & = \frac12 \L^{K L} \dt_{K L} {\mF}_{M N \bm}{}^{\bn} + 2 \mF_{L [N| \bm}{}^{\bn} \dt_{|M] K} \L^{L K} - \frac12 \mF_{M N \bm}{}^{\bn} \partial_{K L}{\Lambda^{K L}}\\
    \d_{\L} \mF_{ABC}{}^{D} & = \mL_{\L} \mF_{ABC}{}^{D} = \frac{1}{2} \L^{K L} {\dt}_{K L}{\mF_{ABC}{}^{D}}.
\end{aligned}
\end{equation}
For the transformation in the third line we have used the result of appendix \ref{gendiffandweight}, where for the zero weight scalar $\phi$ we take $e_{(7)}^{-\frac17} e^{\bn}{}_{\m}$. Together with Table \ref{tab:weightsF} the above can be summarized in Table \ref{tab:weightsF2} listing weights of all fluxes of the theory
\begin{table}[http]
    \centering
    \begin{tabular}{|c|c|c|c|c|c|c|c|}
    \hline
         & $\mF_{\mu \n}{}^{MN}$ & $\mF_{\mu \nu \r M}$ & $\mF_{\m\n\r\l}{}^M$ &  $\mF_{\bm \bn}{}^{\bl}$ & ${\mF^{(E)}}_{\m A}{}^{B}$ & ${\mF}_{M N \bm}{}^{\bn}$ & $\mF_{ABC}{}^{D}$  \\
         \hline
         $\lambda$ & $\frac{1}{5}$ & $\frac45$ & $\frac65$ & $-\frac{1}{5}$ & 0 & $-\frac{1}{5}$ & 0 \\
         \hline
         $\tilde{\lambda}$ & $1$ & $0$ & $1$ & $-\frac{1}{5}$ & 0 & $-1$ & 0  \\
         \hline
    \end{tabular}
    \caption{Weights under generalized Lie derivative. $\l$ is the weight used in formulas (\ref{GenLieY})-(\ref{Liescalar}), and $\tilde{\lambda}$ is an reduced weight - the number standing in front of the term $\dt_{\mK} \L^{\mK}$ in generalized lie derivative, after all symlifications (as in (\ref{LieoftensorsExFT})). It is the weight $\tilde{\lambda}$ that is additive.}
    \label{tab:weightsF2}
\end{table}

\subsection{Flux Lagrangian}

Recall our notations for the generalized vielbeins of the SL(5) ExFT depending on whether the determinant is unity or not:
\begin{equation}\label{VielbeinsRescaling}
            \mE^{A}{}_M \in SL(5),\quad
            E^{A}{}_M = e_{(7)}^{-\frac{1}{14}} \mE^{A}{}_M \in SL(5) \times \mathbb{R}^{+},
\end{equation}
where $e_{(7)} = \det \, e^{\bm}{}_{\m}$ and $E = \det \, E^{A}{}_{M} = e_{(7)}^{-\frac{5}{14}}$. The corresponding generalized metrics read
\begin{equation}\label{GeneralizedMetrics4}
            m_{M N} = m^{A B} \mE^{A}{}_M \mE^B{}_{N},\quad
            M_{M N} = m^{A B} E^{A}{}_M E^{B}{}_N, \quad m_{MN} = e^{\frac17}_{(7)} M_{MN} = M^{-\frac15} M_{MN},
\end{equation}
and we will also use
\begin{equation}\label{GeneralizedMetrics5}
            \mM^{MN, KL} = m^{M K} m^{N L} - m^{M L} m^{N K}, \quad \mM^{M,K} \equiv \mM^{MN,}{}^{K}{}_{N} = 4 m^{M K}.
\end{equation}
It is important to stress out, that although the definition of the vielbein $E^A{}_M$ reminds that of \cite{Gubarev:2020ydf} we do not impose any kind of truncation and consider the full SL(5) exceptional field theory. In particular this means that in (\ref{VielbeinsRescaling}) the vielbein $e^{\bm}{}_{\m}$ depends on the full set of coordinates $(x^{\m}, \mathbb{X}^{[MN]})$.

In the standard formulation the full SL(5) ExFT Lagrangian takes the following form (see \cite{Musaev:2015ces})
\begin{equation}
\label{eq:sl5lagrangian}
    \begin{aligned}
     e_{(7)}^{-1}\mL =& \ \hat{R}[g_{(7)}] \mp \fr18 m_{MN}m_{KL}\mF_{\m\n}{}^{MK}\mF^{\m\n N L}+\frac{1}{4} g^{\m\n}\DD_{\m}m_{M N} \DD_\n m^{M N} \pm e_{(7)}^{-1} V\\
     &+\fr{1}{3\cdot (16)^2}m^{MN} \mF_{\m\n\r M}\mF^{\m\n\r}{}_N + e_{(7)}^{-1} \mL_{top},
    \end{aligned}
\end{equation}
where the term $V$ usually referred to as the potential contains only derivatives w.r.t. the extended coordinates
\begin{equation}\label{potentialV}
\begin{aligned}
 e_{(7)}^{-1} V = & \ \frac{1}{8}\,  {\partial}_{MN}{{m}_{P Q}}\,  {\partial}_{KL}{{m}^{P Q}}\,  {m}^{M K} {m}^{N L}      +\fr12 m^{MK}\dt_{MN}m^{NL}(g_{(7)}^{-1}\dt_{KL}g_{(7)}) \\
 &+ \frac{1}{2}\, {\partial}_{MN}{{m}^{L N}}\,  {\partial}_{KL}{{m}^{M K}} +\frac{1}{2}\, {\partial}_{MN}{{m}_{P Q}}\,  {\partial}_{KL}{{m}^{M P}}\,  {m}^{N K} {m}^{L Q}\\ & + \fr18 m^{MK}m^{NL}(g_{(7)}^{-1}\dt_{MN}g_{(7)})(g_{(7)}^{-1}\dt_{KL}g_{(7)}) + \fr18 m^{MK}m^{NL}\dt_{MN}g^{\m\n}\dt_{KL}g_{\m\n}.
\end{aligned}
\end{equation}
Here and in what folows the upper sign corresponds to the case when the external $d=7$ space is of the Lorentzian signature, and the lower sign stand for the Euclidean signature (certainly we assume that the general signature of the 11-dimensional space is Lorentzian). The case when the time-like direction lies in the internal space of ExFT (the lower sign) is convenient for studies of time-like $U$-dualities. 

To show that the Lagrangian \eqref{eq:sl5lagrangian} can be written completely in terms of generalized fluxes defined in \eqref{sl5fluxes} and \eqref{TYZfluxes} we first focus at the potential term \eqref{potentialV} and notice that the last term is not of the desired form. However, one finds the relation
\begin{equation}
    \dt_{MN} g_{\m\n} = 2 \mF_{M N (\m \n)} + \frac17 \,  g_{\m \n}  g_{(7)}^{-1} \dt_{M N} g_{(7)},
\end{equation}
and similarly for $\dt_{MN}g^{\m\n}$, that follows from contraction of the flux $\mF_{MN\bm}{}^{\bn}$ with external vielbein. Using that we obtain 
\begin{equation}
\begin{aligned}
    & \fr18 m^{MK}m^{NL}\dt_{MN}g^{\m\n}\dt_{KL}g_{\m\n} =\\
    & - \frac12 m^{MK}m^{NL} \mF_{M N (\m \n)} \mF_{KL}{}^{(\m \n)} - \frac{1}{56} \, m^{MK}m^{NL} g_{(7)}^{-2} \dt_{KL} g_{(7)} \dt_{M N} g_{(7)},
\end{aligned}
\end{equation}
where we have used tracelessness of the flux $\mF_{KL}{}^{\m}{} _{\m} = 0$.
Then the potential (\ref{potentialV}) can be written in terms of the rescaled generalized vielbeins $E_{M}{}^{A}$ and the corresponding fluxes $\mF_{ABC}{}^{D}$ (\ref{internalfluxes}) as
\begin{equation}\label{potentialVfluxes}
\begin{aligned}
 e_{(7)}^{-\frac57} V = & - \frac{700}{3}\, {\q}_{A B} {\q}_{C D\, } {m}^{A C} {m}^{B D\, } + {Y}_{A B} {Y}_{C D\, } {m}^{A C} {m}^{B D\, } - \frac{1}{2}\, {Y}_{A B} {Y}_{C D\, } {m}^{A B} {m}^{C D\, } \\
 & + \frac{9}{4}\, {Z}_{A B C}\,^{D\, } {Z}_{D\,  E F}\,^{A} {m}^{B E} {m}^{C F} + \frac{3}{4}\, {Z}_{A A1 B}\,^{C} {Z}_{D\,  E F}\,^{G} {m}_{C G} {m}^{A D\, } {m}^{A1 E} {m}^{B F}\\
 & - \frac{e_{(7)}^{\frac27}}{2} m^{MK}m^{NL} \mF_{M N (\m \n)} \mF_{K L}{}^{(\m \n)}.
\end{aligned}
\end{equation}
The first two line are of the same form as the scalar potential of the $D=7$ maximal gauge supergravity \cite{Weidner:2006rp}, however fluxes are not required to be constants. The last term in \eqref{potentialVfluxes} distinguishes this result from the flux potential used in \cite{Gubarev:2020ydf} and is due to the allowed dependence of the external space-time metric on the coordinates $\XX^{MN}$ on the extended space.

We proceed now with the kinetic term for scalar fields encoded in the generalized metric $m_{MN}$, that can be straightforwardly written as 
\begin{equation}
\begin{aligned}
    e_{(7)}^{-1} \mL_{kin.} & = \frac{1}{4} g^{\m\n}\DD_{\m}m_{M N} \DD_\n m^{M N}  \\
    &=  - \frac{1}{2} (\delta_{L}{}^{M} \delta_{N}{}^{K} + m^{MK} m_{NL}) {\mF}_{\mu M}{}^{N} {\mF}_{\nu K}{}^{L} {g}^{\mu \nu} - \frac{1}{3}\, {\mF}_{\mu M}\,^{M} {\mF}_{\nu N}\,^{N} {g}^{\mu \nu} \\
    & = - \frac{1}{2} (\delta_{L}{}^{M} \delta_{N}{}^{K} + m^{MK} m_{NL}) {\mF}_{\mu M}{}^{N} {\mF}_{\nu K}{}^{L} {g}^{\mu \nu},
\end{aligned}
\end{equation}
where we have used $\mF_{\mu N}\,^{N} = 0$.
Note that here flat SL(5) indices can be turned into curved ones by either rescaled or the original generalized vielbein, i.e.  ${\mF}_{\nu K}{}^{L} = E^{A}{}_{K} E_{B}{}^{L} {\mF}_{\nu A}{}^{B} = \mE^{A}{}_{K} \mE_{B}{}^{L} {\mF}_{\nu A}{}^{B}$.

Finally, we turn to the discussion of the topological Lagrangian $\mL_{top}$. As in the case of maximal gauged supergravity it cannot be expressed in terms of covariant quantities, but its variation can. Hence, in terms of covariantized variations (\ref{covariantvariationsfields}) the latter takes the following form\footnote{Let us mention here the misprint in \cite{Musaev:2015ces}, where the correct coefficient is $A = \frac{1}{16 \cdot (4!)^2}$, rather that  $A = \frac{1}{16 \cdot (4!)}$ in the text. The difference is actually inherited from the earlier typo in the transformation  (4) of (4.16) in the text, where instead of $\frac{1}{16}$ must be $- \frac{1}{16 \cdot 4!}$, that changes (4.18) and gives $A = \frac{1}{16 \cdot (4!)^2}$.}
\begin{equation}
\label{var_top}
 \begin{aligned}
  \d \mc{L}_{top}= \frac{1}{16 \cdot (4!)^2} \e^{\m\n\r\l\s\t\k} \bigg[\mF_{\m\n\r\l}{}^{M}\dt_{MN} \D C_{\s\t\k}^{N} + 6 \mF_{\m\n}{}^{MN} \mF_{\r\l\s M}\D B_{\t\k N} - 2 \mF_{\m\n\r M} \mF_{\l\s\t N} \d A_\k^{MN}\bigg]. 
 \end{aligned}
\end{equation}
Field equations for the (non-dynamical) field $C_{\m\n\r}{}^M$ take the form of a (derivative of a) duality relation between fluxes
\begin{equation}
 \dt_{MK}\big(e_{(7)} m^{MN} \mF^{\m\n\r}{}_{N} - \frac{1}{4!} \e^{\m\n\r\l\s\t\k} \mF_{\l\s\t\k}{}^M)=0.
\end{equation}
Upon taking appropriate solutions of the constraints (\ref{2cond}), these relations reduce to the required first-order duality
\begin{equation}\label{BCduality}
e_{(7)} m^{MN} \mF^{\m\n\r}{}_{N} = \frac{1}{4!} \e^{\m\n\r\l\s\t\k} \mF_{\l\s\t\k}{}^M.
\end{equation}
As a result the Lagrangian of SL(5) ExFT (except topological term) and its EOMs can be rewritten completely in term of fluxes as defined above and takes the following form
\begin{equation}
\label{eq:fullfluxlagrangian}
    \begin{aligned}
     e_{(7)}^{-1}\mL =& \ \hat{R}[g_{(7)}] \mp \fr18 m_{MN}m_{KL}\mF_{\m\n}{}^{MK}\mF^{\m\n N L} +\fr{1}{3\cdot (16)^2}m^{MN} \mF_{\m\n\r M}\mF^{\m\n\r}{}_N + e_{(7)}^{-1} \mL_{top}\\
     & \pm e_{(7)}^{-\frac{2}{7}} \bigg( - \frac{700}{3}\, {\q}_{A B} {\q}_{C D\, } {m}^{A C} {m}^{B D\, } + {Y}_{A B} {Y}_{C D\, } {m}^{A C} {m}^{B D\, } - \frac{1}{2}\, {Y}_{A B} {Y}_{C D\, } {m}^{A B} {m}^{C D\, } \\
    & + \frac{9}{4}\, {Z}_{A B C}\,^{D\, } {Z}_{D\,  E F}\,^{A} {m}^{B E} {m}^{C F} + \frac{3}{4}\, {Z}_{A A1 B}\,^{C} {Z}_{D\,  E F}\,^{G} {m}_{C G} {m}^{A D\, } {m}^{A1 E} {m}^{B F}\\
    & - \frac{e_{(7)}^{\frac27}}{2} m^{MK}m^{NL} \mF_{M N (\m \n)} \mF_{K L}{}^{(\m \n)} \bigg) - \frac{1}{2} (\delta_{L}{}^{M} \delta_{N}{}^{K} + m^{MK} m_{NL}) {\mF}_{\mu M}{}^{N} {\mF}_{\nu K}{}^{L} {g}^{\mu \nu},
    \end{aligned}
\end{equation}

\subsection{Bianchi identities}\label{diffeoBI}

The Lagrangian of the SL(5) exceptional field theory is now written completely in terms of generalized fluxes (and the external metric), that is a collective term for both the scalar fluxes $\mc{F}_{ABC}{}^D$ and field strengths for gauge fields. It is suggestive to recall the standard Maxwell theory with the Lagrangian $\mc{L} = F_2\wedge * F_2$, where $F_2$ is a 2-form and is an example of what we call flux here, rather than the canonical field. Hence, to derive equations of motion one has to introduce gauge potential, that is done using Bianchi identities. When starting without premises one is free to impose either $dF=0$ or $d*F=0$, with the former leading to electric gauge potential, and the latter --- to magnetic. We are here more in the situation when the canonical degrees of freedom are known from the beginning and hence our Bianchi identities will be actual identities, i.e. hold trivially, and in this sense similar to $dF=ddA\equiv 0$. However, adding a source to their RHS one arrives at field equations in the presence of magnetically charged sources (various branes to be discussed in Section \ref{sec:exotic}), or tadpole cancellation conditions when understood as equations on the background field of the Type II  superstring. The full list of Bianchi identities of the SL(5) exceptional field theory is the following
\begin{subequations}
\label{BIfirst}
\begin{align}
Z_{\m\n\r\s\t}{}^M = & -5 \, \DD_{[\m} \mF_{\n \r \s \tau]}{}^{M} +\  \frac{1}{2} \, \epsilon^{M N P K L} \dt_{N P} \mF_{\m \n \r \s \tau K L} - 10 \mF_{[\m \n}{}^{M N} \mF_{\r \sigma \tau] N} =0 , \label{BI0}\\
Z_{\m\n\r}{}^{MN} = & - 3 \, \DD_{[\m} \mF_{\n \r]}{}^{M N}  - \frac1{16} \, \epsilon^{M N P Q R} \dt_{P Q} \mF_{\m \n \r R} = 0,\label{BI1}\\
Z_{\m\n\r\s\, M}  =& - 4 \, \DD_{[\m} \mF_{\n \r \sigma] M}  \, 6 \, \epsilon_{M P Q R S} \mF_{[\m \n}{}^{P Q} \mF_{\r \sigma]}{}^{R S} - \dt_{M N} \mF_{\m \n \r \sigma}{}^{N} = 0,\label{BI2}\\
Z_{\bm \bn \bl }{}^\br  = & - 3 \, \DD_{[\bm} \mF_{\bn \bl]}{}^{\br}  - \frac32 \, \mF_{[\bm \bn|}{}^{M N} \mF_{M N |\bl]}{}^{\br} + \frac35 \, \mF_{\bs [\bm|}{}^{M N} \mF_{M N |\bn}{}^{\bs} \delta_{\bl]}{}^{\br} \nonumber \\
& - \frac3{10} \, e_{(7)}^{-1} \dt_{M N}(e_{(7)} \mF_{[\bm \bn}{}^{M N} \delta_{\bl]}{}^{\br} )+ \frac32 \mF_{[\bm \bn}{}^{\bs} \mF_{\bl] \bs}{}^{\br} = 0, \label{BI3}\\
Z_{\bm\bn} = & - 2 \, \DD_{[\bm} G_{\bn]}  - \frac7{10} \, e_{(7)}^{-1} \dt_{M N}(e_{(7)} \mF_{\bm \bn}{}^{M N}) - \mF_{\bm \bn}{}^{\br} G_{\br} + \frac75 \, \mF_{M N [\bm}{}^{\br} \mF_{\bn] \br}{}^{M N} = 0, \label{BI4}\\
Z_{\bm\bn \, MN}{}^\bl = & -2\, {\DD}_{[\bm|}{{\mF}_{M N |\bn]}\,^{\bl}}\,  - \frac{2}{7}\, {\mF}_{M N [\bm}\,^{\br} {\delta}_{\bn]}\,^{\bl} {G}_{\br} + 2\, {\mF}_{[\bm| \br}\,^{\bl} {\mF}_{M N |\bn]}\,^{\br} - {\mF}_{\bm \bn}\,^{\br} {\mF}_{M N \br}\,^{\bl} \nonumber\\
&\, +  e_{(7)}^{-\frac{1}{7}} {\partial}_{M N}(e_{(7)}^{\frac{1}{7}} {{\mF}_{\bm \bn}\,^{\bl}})\,- \frac{2}{7}\, e_{(7)}^{-\frac{1}{7}} {\partial}_{M N}(e_{(7)}^{\frac{1}{7}} {{G}_{[\bm}})\,  {\delta}_{\bn]}\,^{\bl} =0, \label{BI5}\\
Z_{\m\n \, A}{}^B = & - 2 \DD_{[\m} \mF_{\n] A}{}^{B}  +  {\mF}_{C D\,  A}\,^{B} {\mF}_{\mu \nu}\,^{C D\, } - {\mF}_{\mu \nu}\,^{C D\, } {\delta}_{A}\,^{B} {\theta}_{C D\, } - {\partial}_{A C}{{\mF}_{\mu \nu}\,^{B C}}\, \nonumber \\
& + \frac{1}{5}\, {\delta}_{A}\,^{B} {\partial}_{C D}{{\mF}_{\mu \nu}\,^{C D\, }}\,  - 2 {\mF}_{[\mu| A}\,^{C} {\mF}_{|\nu] C}\,^{B} =0 ,\label{BI6} \\
Z_{\m\, ABC}{}^D = & -\DD_{\m} \mF_{A B C}{}^{E}  - \frac{1}{2}\, {\partial}_{A B}{{\mF}_{\mu C}\,^{E}}\,  + \frac{1}{14}\, {\delta}_{C}\,^{E} {\partial}_{A B}{{G}_{\mu}}\,  - \frac{1}{2}\, {\partial}_{B C}{{\mF}_{\mu A}\,^{E}}+ \frac{1}{14}\, {\delta}_{A}\,^{E} {\partial}_{B C}{{G}_{\mu}}\, \, \nonumber \\
&+ \frac{1}{2}\, {\delta}_{A}\,^{E} {\partial}_{C D\, }{{\mF}_{\mu B}\,^{D\, }}\,  + \frac{1}{4}\, {\delta}_{C}\,^{E} {\partial}_{B D\, }{{\mF}_{\mu A}\,^{D\, }}\,  - \frac{1}{4}\, {\delta}_{C}\,^{E} {\partial}_{A D\, }{{\mF}_{\mu B}\,^{D\, }}\,  \nonumber \\
&- \frac{1}{2}\, {\delta}_{B}\,^{E} {\partial}_{C D\, }{{\mF}_{\mu A}\,^{D\, }}\, - \frac{1}{14}\, {\delta}_{B}\,^{E} {\partial}_{A C}{{G}_{\mu}}   + \frac{1}{2}\, {\partial}_{A C}{{\mF}_{\mu B}\,^{E}}\,  + {\mF}_{\mu A}\,^{D\, } {\mF}_{B D\,  C}\,^{E} \nonumber \\
&- {\mF}_{\mu B}\,^{D\, } {\mF}_{A D\,  C}\,^{E} - {\mF}_{\mu C}\,^{D\, } {\mF}_{A B D\, }\,^{E} + {\mF}_{\mu D\, }\,^{E} {\mF}_{A B C}\,^{D\, }  + \frac{1}{7}\, {\mF}_{A B C}\,^{E} {G}_{\mu} = 0 , \label{BI7}\\
Z_{MN,KL\, \bm}{}^\bn = & \ \partial_{M N}{\mF_{K L \bm}{}^{\bn}} - \partial_{K L}{\mF_{M N \bm}{}^{\bn}} - \mF_{K L \bm}{}^{\bl} \mF_{M N \bl}{}^{\bn} + \mF_{M N \bm}{}^{\bl} \mF_{K L \bl}{}^{\bn} = 0,\label{BI8} \\
Z_{DF,ABC}{}^E = &\ \frac32 \partial_{[A B|}{\mF_{D F |C]}{}^{E}} - \frac12 \partial_{D F}{\mF_{A B C}{}^{E}} + \partial_{C G}{\mF_{D F [A}{}^{G}} \delta_{B]}{}^{E} - \frac14 \delta_{C}{}^{E} \partial_{G [A|}{\mF_{D F |B]}{}^{G}} \nonumber\\
& + 2 \mF_{[A| G C}{}^{E} \mF_{D F |B]}{}^{G} + \mF_{A B G}{}^{E} \mF_{D F C}{}^{G} - \mF_{A B C}{}^{G} \mF_{D F G}{}^{E} = 0 \label{BI9},
\end{align}
\end{subequations}
where we have used the notation $\dt_{AB} = E_{AB}{}^{MN}\dt_{MN}$. The first three Bianchi identitites (enumerated by 0,1 and 2) follow from tensor hierarchy fields (\ref{hierarchy}) and are already known in the literature (see for example \cite{Musaev:2015ces,Berman:2020tqn}). The identities 3, 4 and 5 are straightforward to derive by considering the most general expression of the form
\begin{equation}
    \mD \mF + \text{coefficients} \cdot \mF \mF = 0,
\end{equation}
and then tuning coefficients on the RHS. These are novel to our knowledge. The non-derivative part of the last identity is known under the name of quadratic constraints of the $D=7$ maximal gauged supergravity \cite{Samtleben:2005bp}. A detailed derivation of the above Bianchi identities together with their explicit check can be found in Cadabra files of \cite{sl5flux:git}.

For the general case of non-constant fluxes to derive the identities 6 -- 9 one has to use transformation properties under external and internal diffeomorphisms. In the context of double field theory this approach was advocated in \cite{Geissbuhler:2013uka}. Before turning to exceptional field theory let us demonstrate the trick on the simple case of a U(1) field $A_\m$ and its field strength $F_{\m\n} = 2 \dt_{[\m}A_{\n]}$ (in this calculation small Greek indices a some space-time indices with no relation to the rest of the text). First, one writes transformation of the gauge potential $A_\m$ under a diffeomorphism parametrized by $\x^\m$ back in terms of the flux $F_{\m\n}$:
\begin{equation}
    \d_\x A_\m  = \x^\n \dt_\n A_\m + A_\n \dt_\m \x^\n = \x^\n F_{\n\m} + \dt_\m(\x^\n A_\n).
\end{equation}
The last term above is a gauge transformation and can be dropped in what follows. Now, substitute this into a transformation of the field strength itself to get
\begin{equation}
    \d_\x F_{\m\n} = 2 \dt_{[\m} \d_\x A_{\n]} = L_\x F_{\m\n} + 3 \x^\r \dt_{[\m}F_{\n\r]}.
\end{equation}
Here $L_\x$ is the standard Lie derivative and is the desired transformation of a tensor, and the second term is nothing but the Bianchi identities for the field strength $F_{\m\n}$. Hence, one requires it to vanish, which is trivially the case given the definition of $F_{\m \n}$.

Returning back to exceptional field theory one considers generalized diffeomorphisms acting on the generalized vielbein and rewrites the variation back in terms of fluxes
\begin{equation}
\begin{aligned}
    \delta_{\Lambda} E_{C}{}^{M} = & \frac12 \Lambda^{A B} \partial_{A B}{E_{C}{}^{M}} - E_{C}{}^{L} \partial_{L K}{\Lambda^{M K}}
    + \frac14 E_{C}{}^{M} \partial_{K L}{\Lambda^{K L}} = \\
    = & \mF_{A B C}{}^{E} E_{E}{}^{M} \Lambda^{A B} - E_{A}{}^{M} \partial_{C B}{\Lambda^{A B}} + \frac14 E_{C}{}^{M} \partial_{A B}{\Lambda^{A B}},
\end{aligned}
\end{equation}
and similarly for its inverse. This implies the following transformations for the internal fluxes
\begin{equation}
\begin{aligned}
    \delta_{\Lambda} \mF_{A B C}{}^{E} = \, & \frac12 \Lambda^{D F} \partial_{A B}{\mF_{D F C}{}^{E}} + \frac12 \Lambda^{D F} \partial_{B C}{\mF_{D F A}{}^{E}} - \frac12 \Lambda^{F G} \delta_{A}^{E} \partial_{C D}{\mF_{F G B}{}^{D}}\\
    & - \frac14 \Lambda^{F G} \delta_{C}^{E} \partial_{B D}{\mF_{F G A}{}^{D}} + \frac14 \Lambda^{F G} \delta_{C}^{E} \partial_{A D}{\mF_{F G B}{}^{D}} + \frac12 \Lambda^{F G} \delta_{B}^{E} \partial_{C D}{\mF_{F G A}{}^{D}} \\
    & - \frac12 \Lambda^{D F} \partial_{A C}{\mF_{D F B}{}^{E}} - \mF_{B G C}{}^{E} \mF_{D F A}{}^{G} \Lambda^{D F} + \mF_{A G C}{}^{E} \mF_{D F B}{}^{G} \Lambda^{D F}\\
    & + \mF_{A B G}{}^{E} \mF_{D F C}{}^{G} \Lambda^{D F} - \mF_{A B C}{}^{D} \mF_{F G D}{}^{E} \Lambda^{F G} + \mF_{...} \partial_{...} \Lambda^{...} =\\
    = \, & \frac12 \Lambda^{D F} \partial_{D F}{\mF_{A B C}{}^{E}} + \mZ_{D F, A B C}{}^{E} \Lambda^{D F} \overset{!}{=}  \frac12 \Lambda^{D F} \partial_{D F}{\mF_{A B C}{}^{E}},
\end{aligned}
\end{equation}
where terms with dots are proportional to the section constraint and hence can be dropped. The last equation requires that the flux must transform as a generalized scalar with the appropriate weight, that  results in the Bianchi identities $Z_{DF,ABC}{}^E = 0$ that is \eqref{BI9}. The same flux $\mF_{A B C}{}^{D}$ must also be a scalar under external diffeomorphisms, that gives additional  Bianchi identities. For that we write
\begin{equation}
 \begin{aligned}
    \d_{\xi} E^{A}{}_{M} &= \x^\m \mc{D}_\m E^{A}{}_{M} - \frac1{14} E^{A}{}_{M} \DD_{\m} \x^{\m} = \x^{\m} E^{B}{}_{M} {\mF^{(E)}}_{\m B}{}^{A} - \frac1{14} E^{A}{}_{M} \DD_{\m} \x^{\m} = \\
    &= \x^{\m} E^{B}{}_{M} {\mF_{\m B}{}^{A}} - \frac{1}{14} \x^{\m} E^{A}{}_{M} G_{\m} - \frac1{14} E^{A}{}_{M} \DD_{\m} \x^{\m} ,\\
    \d_{\xi} E_{B}{}^{K} &= - \xi^{\mu} E_{A}{}^{K} \mF_{\nu B}{}^{A} + \frac{1}{14} \xi^{\mu} E_{B}{}^{K} G_{\mu} + \frac{1}{14} E_{B}{}^{K} \mD_{\mu}{\xi^{\mu}}.
 \end{aligned}
\end{equation}
Requiring $\d_\L \mc{F}_{ABC}{}^D  = \x^\m \mc{D}_\m \mc{F}_{ABC}{}^D$ we derive the Bianchi identities \eqref{BI7}. Apparently,  a direct substitution of the flux written in terms of generalized vielbeins fulfills the derived Bianchi identities.

The same procedure is then applied to the fluxes  ${\mF}_{M N \bm}{}^{\bn}$ and  $\mF_{\m A}{}^{B} = {\mF^{(E)}}_{\m A}{}^{B} + \frac{1}{14} \delta_{A}{}^{B} G_{\mu}$. To require covariance of the former under generalized Lie derivative we first write 
\begin{equation}
    \begin{aligned}
    \delta_{\Lambda} e^{\bm}{}_{\m}& = \mL_{\Lambda} e^{\bm}{}_{\m} = \frac12 \L^{K L} \partial_{K L} e^{\bm}{}_{\m} + \frac1{10} e^{\bm}{}_{\m} \partial_{K L} \L^{K L} \\
    &= \frac12 \L^{K L} \bigg(e^{\bn}{}_{\m} \mF^{(e)}_{MN \bn}{}^{\bm} + \frac1{7} e^{\bm}{}_{\m} \mF_{MN}\bigg) + \frac1{10} e^{\bm}{}_{\m} \partial_{K L} \L^{K L},
    \end{aligned}
\end{equation}
that implies
\begin{equation}
    \delta_{\Lambda} {\mF}_{M N \bm}{}^{\bn} = \mL_{\Lambda} {\mF}_{M N \bm}{}^{\bn}+\frac{1}{2}\, {\Lambda}^{K L} Z_{KL,MN\, \bar{\m}}{}^{\bn}. 
\end{equation}
Requiring covariance under external diffeomorphisms we will get the fifth Bianchi identity \eqref{BI5}, that is an additional consistency check.

Covariance of $\mF_{\m A}{}^{B} $ under generalized Lie derivative gives the (already derived) seventh Bianchi identity \eqref{BI7}. For external diffeomorphisms we write 
\begin{equation}
 \begin{aligned}
  \d_{\xi} e^{\bm}{}_\m &= \x^\n \DD_\n e^{\bm}{}_\m + e^{\bm}{}_\n \DD_\m \x^\n ,\\
   \d_{\xi} A_\m{}^{MN} &= \x^\n \mF_{\n\m}{}^{MN} + \frac12 \M^{MN,KL} g_{\m\n} \dt_{KL} \x^\n.\\
 \end{aligned}
\end{equation}
Substituting these together with $\d_{\xi} E^{A}{}_{M}$, $\d_{\xi} E_{B}{}^{K}$ written above in variation of the flux we arrive at
\begin{equation}
    \delta_{\x} \mF_{\n A}{}^{B} = \xi^{\m} \mD_{\m} \mF_{\n A}{}^{B} + \mF_{\m A}{}^{B} \mD_{\n} \xi^{\m}+ \xi^{\mu} Z_{\mu\nu \, A}{}^B ,
\end{equation}
that gives the sixth Bianchi identity  \eqref{BI6}.

\subsection{From curved to flat indices and back again}

Bianchi identities presented in the previous section naturally come with mixed indices: some are written in flat indices, some are written in curved indices. For examples, the first Bianchi identity comes from the standard formulation of exceptional field theory and is understood as a relation between field strengths. In contrast, the seventh identity, that is for generalized anholonomy coefficients, is more naturally written in flat indices since in this case the flux is identified with components of the embedding tensor upon a Scherk-Schwarz reduction. Let us now think of all $\mc{F}'s$ as fluxes and components of some larger (infinitely component) embedding tensor and write all Bianchi identities in flat indices. In this case the NS-NS sector of the reduction SL(5)$\to $GL(4) will reproduce the split form of the Bianchi identities of DFT (see e.g. \cite{Musaev:2019yfn}).

We collect all calculational details in Appendix \ref{app:BIflat} and here we only stress few subtle points. First, the transition to flat indices is performed by contraction with vielbeins as follows
\begin{equation}
\begin{aligned}
    T^{\bm} =&\ e^{\bm}{}_{\m} T^{\m},\\
    T^{A} =&\ E^{A}{}_{M} T^{M},
\end{aligned}
\end{equation}
where for flat flat SL(5) indices we use the generalized vielbein $E^{A}{}_{B} \in SL(5) \times \mathbb{R}^{+}$. This should not cause confusion since all the fluxes in (\ref{sl5fluxes}) and (\ref{sl5fluxesauxil}) have curved SL(5) indices except $\mF_{AB,C}{}^{D}$, that is already written in flat indices using $E^{A}{}_{B}$, and ${\mF^{(E)}}_{\m A}{}^{B}$ and $\mF_{\m A}{}^{B}$ for which we have
\begin{equation}
\begin{aligned}
    {\mF^{(E)}}_{\m A}{}^{B} =&\ E^{A}{}_{K} E_{B}{}^{L} {\mF}_{\nu A}{}^{B} = \mE^{A}{}_{K} \mE_{B}{}^{L} {\mF^{(E)}}_{\nu A}{}^{B},\\
    {\mF}_{\nu K}{}^{L} =&\ E^{A}{}_{K} E_{B}{}^{L} {\mF}_{\nu A}{}^{B} = \mE^{A}{}_{K} \mE_{B}{}^{L} {\mF}_{\nu A}{}^{B}.
\end{aligned}
\end{equation}

Second, the fourth and the sixth Bianchi identities \eqref{BI6} can be rewritten as a single for the flux ${\mF^{(E)}}_{\nu A}\,^{B}$ as follows
\begin{equation}\label{BI6star}
    \tilde{Z}_{\m\n\, A}{}^B = -2 {\DD}_{[\mu}{{\mF^{(E)}}_{\nu] A}\,^{B}} +{\mF}_{C D\,  A}\,^{B} {\mF}_{\mu \nu}\,^{C D\, } - {\partial}_{A C}{{\mF}_{\mu \nu}\,^{B C}}\,  - 2 {\mF^{(E)}}_{[\mu| A}\,^{C} {\mF^{(E)}}_{|\nu] C}\,^{B} =0,
\end{equation}
whose traceless part gives \eqref{BI6} and the trace part reproduces \eqref{BI4} (see the files \texttt{Check of BI 4} and \texttt{Check of BI 6} of \cite{sl5flux:git}). Hence, using the flux ${\mF^{(E)}}_{\nu A}\,^{B}$ two Bianchi identities can be written in a combined form.

We finally present the full set of  Bianchi identities in flat indices
\begin{subequations}
\label{BIflat}
\begin{align}
Z_{\bm\bn\br}{}^{AB} & = - 3 \, \DD_{[\bm} \mF_{\bn \br]}{}^{A B}  - 3 \, \mF_{[\bm \bn|}{}^{\bl} \mF_{\bl | \br]}{}^{A B} - 6 \, \mF^{(E)}{}_{[\bm| C}{}^{[A} \mF_{|\bn \br]}{}^{B] C}, \nonumber \\
&   - \frac1{16} e_{(7)}^{-\frac6{14}} \, \epsilon^{A B C D E} \dt_{C D}(e_{(7)}^{\frac1{14}} \mF_{\bm \bn \br E}) - \frac3{16} e_{(7)}^{-\frac5{14}} \, \epsilon^{A B C D E} \mF_{C D [\bm|}{}^{\bl} \mF_{\bl | \bn \br] E} \nonumber \\
& - \frac1{16} e_{(7)}^{-\frac5{14}} \, \epsilon^{A B C D E} \mF_{\bm \bn \br F} \left(\frac{20}{3} \theta_{C D} \delta_{E}{}^{F} - Z_{C D E}{}^{F} \right) = 0 , \label{BIf1}\\
Z_{\bm\bn\br\bs \, A} = & -4 \, \DD_{[\bm} \mF_{\bn \br \bar{\sigma}] A}   - 6 \, \mF_{[\bm \bn|}{}^{\bl} \mF_{\bl | \br \bar{\sigma}] A} - 4 \, \mF^{(E)}{}_{[\bm |A}{}^{C} \mF_{|\bn \br \bar{\sigma}] C} + 6 \, e_{(7)}^{\frac5{14}} \epsilon_{A C D E F} \mF_{[\bm \bn}{}^{C D} \mF_{\br \bar{\sigma}]}{}^{E F}\nonumber\\
& - 4 \, \mF_{A B [\bm|}{}^{\bl} \mF_{\bl | \bn \br \bar{\sigma}]}{}^{B}  - e_{(7)}^{-\frac3{14}} \dt_{A B}(e_{(7)}^{\frac3{14}}  \mF_{\bm \bn \br \bar{\sigma}}{}^{B})- \mF_{\bm \bn \br \bar{\sigma}}{}^{B} \left(2 \mF_{A B D}{}^{D} - \frac12 \mF_{D (A B)}{}^{D}\right),\label{BIf2}\\
Z_{\bm\bn\bl}{}^\br = & -3 \, \DD_{[\bm} \mF_{\bn \bl]}{}^{\br}  - \frac32 \, \mF_{[\bm \bn|}{}^{A B} \mF_{A B |\bl]}{}^{\br} + \frac35 \, \mF_{\bs [\bm|}{}^{A B} \mF_{A B |\bn}{}^{\bs} \delta_{\bl]}{}^{\br} - \frac65 \, \mF_{A B C}{}^{C} \mF_{[\bm \bn}{}^{A B} \delta_{\bl]}{}^{\br}\nonumber\\
& - \frac3{10} \, e_{(7)}^{-\frac27} \dt_{A B}(e_{(7)}^{\frac27} \mF_{[\bm \bn}{}^{A B} ) \delta_{\bl]}{}^{\br}+ \frac32 \mF_{[\bm \bn}{}^{\bs} \mF_{\bl] \bs}{}^{\br} ,\label{BIf3}\\
Z_{AB\, \bm \bn}{}^\bl = & - 2\, {\DD}_{[\bm|}{{\mF}_{A B |\bn]}\,^{\bl}}\,  + \frac{4}{5}\, {\mF}_{A B [\bm}\,^{\br} {\delta}_{\bn]}\,^{\bl} {\mF^{(E)}}_{\br C}\,^{C} + 2\, {\mF}_{[\bm| \br}\,^{\bl} {\mF}_{A B |\bn]}\,^{\br} - {\mF}_{\bm \bn}\,^{\br} {\mF}_{A B \br}\,^{\bl}\nonumber\\
&- 2 {\mF}_{[A| C \bn}\,^{\bl} \mF^{(E)}{}_{\bm |B]}{}^{C} + 2 {\mF}_{[A| C \bm}\,^{\bl} \mF^{(E)}{}_{\bn |B]}{}^{C} ,\nonumber \\
& + \frac{4}{5}\, e_{(7)}^{-\frac17} {\partial}_{A B}{(e_{(7)}^{\frac17} {\mF^{(E)}}_{[\bm| C}\,^{C})}\,  {\delta}_{|\bn]}\,^{\bl}  + e_{(7)}^{-\frac17} {\partial}_{A B}{( e_{(7)}^{\frac17} {\mF}_{\bm \bn}\,^{\bl})}\, \label{BIf5}\\
Z_{\bm\bn\, A}{}^B =& - 2 {\DD}_{[\bm}{{\mF^{(E)}}_{\bn] A}\,^{B}} - \mF_{\bm \bn}{}^{\br} {\mF^{(E)}}_{\br A}\,^{B} + {\mF}_{C D\,  A}\,^{B} {\mF}_{\bm \bn}\,^{C D\, } - e_{(7)}^{-\frac27} {\partial}_{A C}{(e_{(7)}^{\frac27} {\mF}_{\bm \bn}\,^{B C})}\, \nonumber\\
& - 2 {\mF}_{[\bm| \br}\,^{B C} \mF_{A C |\bn]}{}^{\br} - 2 {\mF^{(E)}}_{[\bm| A}\,^{C} {\mF^{(E)}}_{|\bn] C}\,^{B}\, \label{BIf6s} \\
Z_{\bm\, ABC}{}^E = & -\DD_{\bm} \mF_{A B C}{}^{E} - \frac{1}{2}\, e_{(7)}^{- \frac17} {\partial}_{[A B|}{(e_{(7)}^{\frac17} {\mF^{(E)}}_{\bm |C]}\,^{E})}\,  + e_{(7)}^{-\frac17} {\delta}_{[A|}\,^{E} {\partial}_{C D}{(e_{(7)}^{\frac17} {\mF^{(E)}}_{\bm |B]}\,^{D\, })}\, \nonumber \\
& - \frac{1}{2}\, e_{(7)}^{-\frac17} {\delta}_{C}\,^{E} {\partial}_{[A| D}{(e_{(7)}^{\frac17} {\mF^{(E)}}_{\bm |B]}\,^{D\, })}\, \nonumber \\
& - \frac{1}{2}\, {\mF}_{[A B| \bm}{}^{\bn}{{\mF^{(E)}}_{\bn |C]}\,^{E}}\,  + {\delta}_{[A|}\,^{E} {\mF}_{C D \bm}{}^{\bn}{{\mF^{(E)}}_{\bn |B]}\,^{D\, }}\,  - \frac{1}{2}\, {\delta}_{C}\,^{E} {\mF}_{[A| D \bm}{}^{\bn}{{\mF^{(E)}}_{\mu |B]}\,^{D\, }}\,  \nonumber\\
& - 2 {\mF}_{[A| D\,  C}\,^{E} {\mF^{(E)}}_{\bm |B]}\,^{D\, } - {\mF}_{A B D\, }\,^{E} {\mF^{(E)}}_{\bm C}\,^{D\, } + {\mF}_{A B C}\,^{D\, } {\mF^{(E)}}_{\bm D\, }\,^{E}, \label{BIf7}\\
Z_{CD,AB \, \bm}{}^\bn = & \ \partial_{C D}{\mF_{A B \bm}{}^{\bn}} - \partial_{A B} \mF_{C D \bm}{}^{\bn} - 2 \, \mF_{[A| E \bm}{}^{\bn} \mF_{C D, |B]}{}^{E} + 2 \, \mF_{[C| E \bm}{}^{\bn} \mF_{A B, |D]}{}^{E}\nonumber \\
& - \mF_{A B \bm}{}^{\bl} \mF_{C D \bl}{}^{\bn} + \mF_{C D \bm}{}^{\bl} \mF_{A B \bl}{}^{\bn}, \label{BIf8}\\
Z_{DF,ABC}{}^E =& \ \frac32 \partial_{[A B|}{\mF_{D F |C]}{}^{E}} - \frac12 \partial_{D F}{\mF_{A B C}{}^{E}} + \partial_{C G}{\mF_{D F [A}{}^{G}} \delta_{B]}{}^{E} - \frac14 \delta_{C}{}^{E} \partial_{G [A|}{\mF_{D F |B]}{}^{G}} \nonumber\\
& + 2 \mF_{[A| G C}{}^{E} \mF_{D F |B]}{}^{G} + \mF_{A B G}{}^{E} \mF_{D F C}{}^{G} - \mF_{A B C}{}^{G} \mF_{D F G}{}^{E}. \label{BIf9}
\end{align}
\end{subequations}

\section{Exotic potentials}
\label{sec:exotic}

Before turning to exotic potentials of 11-dimensional supergravity collected into irreps of the SL(5) duality group, recall the simple case of Maxwell theory. Its Bianchi identities $dF = 0$ can be implemented on the level of the action as
\begin{equation}
    S_{EM} = \int d^4x \left( - \fr14 F_{\m\n}F^{\m\n} + \fr12\e^{\m\n\r\s}\tilde{A}_{\m}\dt_{\n}F_{\r\s}\right),
\end{equation}
where $A^{\m\n\r}$ is totally antisymmetric. As before small Greek indices run $\m,\n=0,1,2,3$ only in reference to calculations in the Maxwell theory. Varying w.r.t. the Lagrange multiplier $\tilde{A}_\m$ one obtains the Bianchi identities $\dt_{[\m}F_{\n\r]}=0$ and is able to introduce the standard gauge potential $F_{\m\n} = 2\dt_{[\m}A_{\n]}$. Alternatively, variation w.r.t. to $F_{\m\n}$ relates the field $\tilde{A}_\m$ to the dual field strength
\begin{equation}
    \fr12 \e_{\m\n\r\s}F^{\r\s} = \tilde{F}_{\m\n} = 2 \dt_{[\m} \tilde{A}_{\n]}.
\end{equation}
Hence, the Lagrange multiplier imposing Bianchi identities is interpreted as the potential magnetic dual to the canonical gauge field. Apparently, Bianchi identities for the field $F_{\m\n}$ are field equations for the field $\tilde{A}_\m$.

At the level linear in a metric fluctuation about the flat space-time the same procedure can be repeated for the graviton field, whose field strength will be the anholonomy coefficients. While Maxwell theory in the absence of external sources can be equivalently rewritten in terms of the magnetic dual gauge potential, the same is not true for the gravitational theory, that is known as the dual graviton problem (see though \cite{Hohm:2018qhd} for comments on whether to call this a problem).

\subsection{Dual formulation of Double Field Theory}

As in the case of YM theory Bianchi identities can be implemented into the full DFT action in flux formulation as Lagrange multipliers \cite{Bergshoeff:2016ncb}
\begin{equation}
    S = \int d^{20} \XX e^{-2d} (\mF \mF) + S_{ABCD}D^{ABCD} + S_{AB}D^{AB}.
\end{equation}
Equations of motion for $D^{ABCD}$ and $D^{AB}$, that are the Bianchi identities, render generalized fluxes $\mF_{ABC}$ and $\mF_{A}$ to be of the form \eqref{eq:fluxesDFT}. Even though there is no proper formulation of DFT in terms of a functional integral, we will refer to this process as ``integrating out fields'' for short. Hence, integrating out the fields $D^{ABCD}$ and $D^{AB}$ one arrives at the standard formulation of double field theory in terms of supergravity degrees of freedom. Alternatively, one might try to integrate out fluxes $\mF_{ABC}$ and $\F_A$ to formulate the theory in terms of $D^{ABCD}$. However, as it has been shown in \cite{Bergshoeff:2016ncb} this does not seem to work out as straightforward as in say Maxwell theory. Note, that on principle that has little to do with non-linearity of the DFT action itself as for example in Type II supergravity one can do so for the 2-form Kalb-Ramond field. It is better to refer this to metric degrees of freedom, as the same behavior is observed in general relativity, where one cannot turn to full non-linear theory of the dual graviton. 

Nonetheless, the magnetic potentials $D^{ABCD}$ and $D^{AB}$ can be understood as fields electrically interacting with certain branes, both standard and exotic, at least at the linear level. Let us illustrate this by comparing to NS-NS potentials interacting with standard and exotic branes taking into account the wrapping rules of \cite{Bergshoeff:2011zk,Bergshoeff:2011ee}. Since these rules are formulated for a D-dimensional maximal supergravity, we should turn to split form of double field theory, where O(10,10) is broken to GL(D)$\times$O(d,d), where $D+d=10$. Without going into too many technical details of the decomposition itself, that can be found e.g. in \cite{Musaev:2019yfn}, let us list the only non-trivial components of the full O(10,10) Bianchi identities upon the decomposition $V^\mA = (V^\a,V_\a,V^A)$:
\begin{equation}
    \label{eq:BI_splitdft}
\begin{aligned}
& S_{A B C D},&& S_{\a A B C}, && S_{\alpha \beta A B}, && S_{\alpha \beta \gamma A},&&
{S}_{\alpha \beta \gamma \delta}, &&S_{A B \alpha }{}^{\beta},  &&
{S}_{\alpha \beta A}{}^{\gamma},&&
S_{\alpha \beta \gamma}{}^{\delta}, &&
S_{\alpha \beta}{}^{\gamma \delta} .
\end{aligned}
\end{equation}
Let us for concreteness assume $D=6$, that allows us to use Hodge star in external six-dimensional space-time to write the following list of magnetic potentials:
\begin{equation}
    \label{eq:pots_DFT}
    \begin{aligned}
       & D_{(6),A B C D},&& D_{(5), A B C}, && D_{(4), A B}, && D_{(3), A},&&
{D}_{(2)}, &&D_{(5,1), A B},  &&
D_{(4,2), A},&&
D_{(3,1)}, &&
D_{(4,2)}.
    \end{aligned}
\end{equation}
Here numbers in parentheses denote: $p$-forms, in case of a single number; potentials of mixed symmetry, in case of two numbers, i.e. a potential $A_{(p,q)}$ has the following components
\begin{equation}
    A_{\a_1\dots\a_p, \b_1\dots \b_q}.
\end{equation}
The above is antisymmetric in the first $p$ and in the latter $q$ entries separately. According to the brane counting rules of \cite{Kleinschmidt:2011vu} only those components for which the indices $[\b_1\dots\b_q]$ equal to $q$ indices of $[\a_1\dots \a_p]$ interact with supersymmetric branes.

\begin{table}[http]
    \centering
    \begin{tabular}{cccccc|cccc|l}
          0 & 1 & 2 & 3 & 4 & 5 &  6 & 7 & 8 & 9 & \\
          \hline
\tm&\tm&   &   &   &   & \tm&\tm&\tm&\tm& $D_{(2)}$ \\
\tm&\tm&  \tm & \od   &   &   & \tm&\tm&\tm&\tm& $D_{(3,1)}$  \\
\tm&\tm& \tm & \tm  & \od  & \od   &   \tm&\tm&\tm&\tm& $D_{(4,2)}$  \\
\tm&\tm&\tm&   &   &   &    &\tm&\tm&\tm& $D_{(3), A}$\\
\tm&\tm&\tm&\tm&   &   &    &   &\tm&\tm& $D_{(4), AB}$\\
\tm&\tm&\tm&\tm&\tm&   &    &   &   &\tm& $D_{(5), ABC}$,\\
\tm&\tm&\tm&\tm&\tm&\tm&    &   &   &   & $D_{(6), ABCD}$ \\
\tm&\tm& \tm  &  \tm &  \tm &  \od &  & &\tm&\tm& $D_{(5,1),AB}$  \\
\tm&\tm& \tm  &\tm   & \od  & \od  &  &\tm&\tm&\tm& $D_{(4,2)A}$  \\
    \end{tabular}
    \caption{Magnetic potentials of DFT according to the Bianchi identities and the corresponding brane alignments. Here $\times$ denotes the worldvolume directions, empty space denotes transverse directions, dotted circle denotes special cycle. The directions $\{6,7,8,9\}$ are doubled, i.e. empty space can be either simply transverse or special. }
    \label{tab:tabdft}
\end{table}

We see, that the first five potentials of \eqref{eq:pots_DFT} correspond to $p$-forms interacting with supersymmetric NS-NS 5-branes, that follow from the analysis of wrapping rules of \cite{Bergshoeff:2011zk,Bergshoeff:2011ee}. The remaining are not covered by their analysis, however follow the same pattern. For example, while $D_{(2)}$ interacts with the NS5-brane wrapping all cycles of the 4-torus, the potentials $D_{(3,1)}$ and $D_{(4,2)}$ interact with the KK5-monopole and $5_2^2$-brane, whose special cycles are in the external space. Note, that we can not see $5_2^3$ and $5_2^4$-branes with special cycles along the external space since these generate fluxes via either the DFT bivector $\b^{mn}$ or dependence on dual coordinates. Either of these do not present in the external space, however, can be organized in the internal space. The corresponding potentials will be $D_{(5,1),A}$ and $D_{4,2},A$, that interact with branes with at one or two special cycles in the external space respectively, and the rest in the internal space. An example of such state can be the $5_2^4$ brane, whose two cycles wrap the internal directions and two wrap the external directions. 

NS 5-branes alignments corresponding to the recovered potentials are listed in Table \ref{tab:tabdft}. Let us illustrate reading of the table by the potentials $D_{(3),A}$ and $D_{(4,2)}$. For the first we have the following components
\begin{equation}
    D_{(3),A}:
    \begin{aligned}
        D_{\m_1\m_2\m_3,m}   &\to D_{\m_1\m_2\m_3 m_1m_2m_3m_4,m}, \\
        D_{\m_1\m_2\m_3}{}^m &\to D_{\m_1\m_2\m_3 m_1m_2m_3}, 
    \end{aligned}
\end{equation}
where $\to$ denote Hodge dualization in four internal directions. Apparently, the potential $D_{\m_1\m_2\m_3 m_1m_2m_3}$ in the second line is a component of the 6-form field interacting with NS5-brane in 10d, while the potential $D_{\m_1\m_2\m_3 m_1m_2m_3m_4,m}$ is that for the KK5-monopole. Hence, we conclude that the former interacts with the NS5-brane whose three worldvolume directions are along the internal space, while the latter interact with the KK5-monopole, whose three world-volume directions and the special Taub-NUT cycle are in the internal space. Note, that Hodge dualization automatically assures that the index $m$ after the comma must be equal to one of the four internal indices before the comma.
\begin{table}[http]
    \centering
    \begin{tabular}{cccccc|cccc|cl}
          0 & 1 & 2 & 3 & 4 & 5 &  6 & 7 & 8 & 9 &  Branes &\\
          \hline
\tm&\tm&  &  & \od  & \od   &  \tm & \tm &\tm&\tm& $5_2^2$ & $D_{(4,2)}$  \\
\tm&\tm&\tm&   &   &   &    &\tm&\tm&\tm& NS5 &\multirow{2}{*}{$D_{(3), A}$} \\
\tm&\tm&\tm&   &   &   &  \od   &\tm&\tm&\tm& KK5 &
    \end{tabular}
    \caption{Part of Table \ref{tab:tabdft} with representations of O(4,4) expanded w.r.t its GL(4) subgroup}
    \label{tab:tabdftpart}
\end{table}

Next, upon Hodge dualization of the $D_{(4,2)}$ we have in component notations 
\begin{equation}
    D_{\m_1\m_2\m_3\m_4 m_1m_2m_3m_4, \n_1\n_2},    
\end{equation}
that is part of the (8,2) mixed symmetry potential in 10d interacting with $5_2^2$-brane. According to the brane wrapping rules, the indices $\n_1\n_2$ must be equal to two of $[\m_1\dots\m_4]$. The remaining two label form components interacting with two world-volume directions of the $5_2^2$-brane in the external space-time. We illustrate this on Table \ref{tab:tabdftpart}.

\subsection{Exotic potentials in D=7 supergravity}

Let us list the magnetic potentials associated with Bianchi identities of the SL(5) theory derived above:
\begin{equation}
    \label{eq:forms7}
    \begin{aligned}
       & A_{(3),\mathbf{5}}, && A_{(4), \ol{\mathbf{10}}}, &&  A_{(5),\mathbf{24}+ \mathbf{1}}, && A_{(6),\mathbf{\ol{10} + \ol{15} + \ol{40}}}, && A_{(7),\mathbf{5+\ol{45}+\ol{70}}}\\
        &A_{(4,1),\mathbf{1}}, && A_{(5,1),\mathbf{10}}, && A_{(6,1),\mathbf{45}},
    \end{aligned}
\end{equation}
where as before numbers in parentheses denote rank or mixed symmetry of a potential, irreps of SL(5) are given in bold. To compare this with the supersymmetric brane counting following from the E$_{11}$ decomposition let us copy the $D=7$ row of Table 1 of \cite{Kleinschmidt:2011vu}.
\begin{table}[http]
    \centering
    \begin{tabular}{|c|c|c|c|c|c|c|c|}
    \hline
         $p$   & 1  & 2 & 3 & 4 & 5 & 6 & 7\\
         \hline
         $D=7$ & $\ol{\mathbf{10}}$ & $\mathbf{5}$ & $\ol{\mathbf{5}}$ & ${\mathbf{10}}$ & ${\mathbf{24}}$ & $\mathbf{40} + \ol{\mathbf{15}}$ & $\mathbf{5} + \ol{\mathbf{45}} + \ol{\mathbf{70}}$ \\
         \hline
    \end{tabular}
    \caption{The tensor hierarchy of $p$-forms for $D=7$ supergravity as predicted by E$_{11}$.}
    \label{tab:axel1}
\end{table}

We see that the set of forms following from tensor hierarchy as presented in Table \ref{tab:axel1} and the set of forms \eqref{eq:forms7} following from Bianchi identities have intersection, but do not fully coincide. Let us give comments on how one should understand this. Start with the second line of \eqref{eq:forms7}, that lists potentials of mixed symmetry. These are not covered by the brane counting approach of \cite{Kleinschmidt:2011vu}, as it focuses on $p$-forms, equivalently speaking on BPS-states corresponding to branes with all special cycles placed along internal directions. 

The 2-form potential in the $\mathbf{5}$ seen from tensor hierarchy is nothing but the 2-form $B_{\m\n M}$ and is magnetically dual to $A_{\m\n\r}{}^M$, that is denoted as $C_{\m\n\r}{}^M$ in the exceptional field construction of \cite{Musaev:2015ces}. At the ExFT level the corresponding duality relation follows from equations of motion for the $3$-form potential. Given that Bianchi identities for the 4-form field strength $\mF_{\m\n\r\s}{}^M$ are simply equations of motion for the field $B_{\m\n M}$. The same is true for the 1-form potential in the $\ol{\mathbf{10}}$, that is magnetically dual to the vector field $A_\m{}^{MN}$. The former however does not appear in  exceptional field theory in contrast to $C_{\m\n\r }{}^M$, that is needed to write Chern-Simons terms in the action and gauge transformations of the physical fields.

Finally, the potentials given by the singlet 5-form and by the 6-form in the $\ol{\mathbf{10}}$ correspond to imaginary roots in the E${}_{11}$ decomposition and are not usually included to the brane counting analysis. The reason is that these are conjectured to not interact with supersymmetric branes. Hence, modding out by the magnetic duality and counting of imaginary root and mixed symmetry potentials the field content of Table \ref{tab:axel1} is the same as the one recovered from Bianchi identities, as it should be.

Let us now give more details on supersymmetric branes these potentials interact with, which in particular will give additional confirmation of the electric-magnetic equivalence between 1-, 2- and 3-,4-form potentials.
\begin{table}[http]
    \centering
    \begin{tabular}{c|ccccccc|cccc|c|c}
         $p$& 0 & 1 & 2 & 3 & 4 & 5 & 6 & 7 & 8 & 9 & 10 & w$\#$ & brane  \\
         \hline
         \multirow{2}{*}{0}& \tm &  &  & & & & & \tm & \tm &  & & 6 & M2\\
          & \tm &  &  & & & & & * & & & & 4 & PP\\
          \hline
          \multirow{2}{*}{1}& \tm & \tm &  & & & & & \tm & \tm & \tm & \tm & 1 & M5\\
          & \tm & \tm &  & & & & &\tm & & & & 4 & M2\\
         \hline
         \multirow{2}{*}{2}& \tm & \tm & \tm & & & & & & & & & 1 & M2\\
         & \tm & \tm & \tm & & & & & \tm & \tm & \tm & & 4 & M5\\
          \hline
         \multirow{2}{*}{3}& \tm & \tm & \tm & \tm & & & & \tm & \tm &  & & 6 & M5\\
          & \tm & \tm & \tm & \tm & & & & \tm &\tm &\tm  &\od & 4 & KK6\\ 
         \hline
         \multirow{3}{*}{4}& \tm & \tm & \tm &  \tm&  \tm& & & \tm & \
          &   & & 4 & M5\\
          & \tm & \tm & \tm & \tm &  \tm& & & \tm & \tm & \od & & 12 & KK6\\
          & \tm & \tm & \tm & \tm & \tm & & & \tm & \od & \od & \od &  4 & $5^3$\\
          \hline
           \multirow{5}{*}{5}& \tm & \tm & \tm &  \tm&  \tm&\tm & &  &  &   & & 1 & M5\\
          & \tm & \tm & \tm & \tm &  \tm&\tm & & \tm & \od &  & & 12 & KK6\\
          & \tm & \tm & \tm & \tm & \tm &\tm & & \od & \od & \od & & 4 & $5^3$\\
          & \tm & \tm & \tm & \tm & \tm &\tm & & \od & \od & \od & $\otimes$& 4 & $5^{(1,3)}$\\
          & \tm & \tm & \tm & \tm & \tm &\tm & & \tm & \tm & \tm & $\otimes$& 4 & $8^{(1,0)}$\\
          \hline
         \multirow{3}{*}{6}& \tm & \tm & \tm &  \tm&  \tm&\tm &\tm & \od &  &   & & 4 & KK6\\
          & \tm & \tm & \tm & \tm &  \tm&\tm & \tm& \tm & \tm & $\otimes$ & & 12 & $8^{(1,0)}$\\
          & \tm & \tm & \tm & \tm & \tm &\tm &\tm & $\otimes$ & $\otimes$ & $\otimes$ & \od& 4 & $6^{(3,1)}$\\
          \hline
    \end{tabular}
    \caption{Brane embeddings corresponding to $p$-form potentials, total winding number counting w$\#$ for each $p$ corresponding to the number of BPS states as count in \cite{Kleinschmidt:2011vu}. Here \tm{} denotes world-volume directions,  \od{} denotes quadratic special cycle, $\otimes${} denotes cubic special cycle, star $*$ denotes direction of the momentum.}
    \label{tab:normal7}
\end{table}
All possible embeddings of M-theory branes both standard and exotic with all special cycles along the internal space are collected in Table \ref{tab:normal7}. Given the four-dimensional internal space is a torus, their windings give counting of BPS states as found in \cite{Kleinschmidt:2011vu}. Note, that in general this counting does not equal to dimension of an SL(5) irrep the potential belongs to. Say, the 4-form potential belongs to the $\mathbf{24}$ of SL(5) and interacts with only 20 brane states. The simplest case where one encounters such a behavior is the doublet of $(p,q)$ 7-branes of Type IIB theory, interacting with a potential transforming as a triplet of the S-duality group (see e.g. \cite{Bergshoeff:2010xc} for more detailed discussion).

The first four rows of Table \ref{tab:normal7} describing brane embeddings for $p=0,1,2,3$ show the electric-magnetic symmetry discussed above. Indeed, upon the EM duality one interchanges M2 with M5, and KK6 with the pp-wave states. That effectively switches $p=1$ with $p=2$, and $p=0$ with $p=3$, as we have already observed at the level of potentials. 

Embeddings of (exotic) branes producing states that interact with mixed symmetry potentials listed in \eqref{eq:forms7} are given in Table \ref{tab:exotic7}.
\begin{table}[http]
    \centering
    \begin{tabular}{c|ccccccc|cccc|c|c}
          potential& 0 & 1 & 2 & 3 & 4 & 5 & 6 & 7 & 8 & 9 & 10 & w$\#$ & brane  \\
         \hline
         (4,1)& \tm & \tm & \tm & \od & & & & \tm & \tm & \tm &\tm & 1 & KK6\\
         \hline
          \multirow{2}{*}{(5,1)}& \tm & \tm & \tm & \tm & \od & & & \tm & \tm & \tm & & 4 & KK6\\
          & \tm & \tm & \tm & \tm & \od & & & \tm & \tm & \od & \od & 6 & $5^3$\\
          \hline
          \multirow{3}{*}{(6,1)}& \tm & \tm & \tm &  \tm&  \tm & \od & & \tm & \tm 
          &   & & 6 & KK6\\
          & \tm & \tm & \tm &  \tm&  \tm&  \od&  & \tm&  \od & \od
          &   & 12 & $5^3$\\
          & \tm & \tm & \tm &  \tm&  \tm&  \od & & \tm&  \od & \od
          &  $\otimes$ & 12 & $5^{(1,3)}$\\
          \hline
          \multirow{4}{*}{(7,1)}& \tm & \tm & \tm &  \tm&  \tm& \tm & \od & \tm & \
          &   & & 4 & KK6\\
          & \tm & \tm & \tm &  \tm&  \tm& \tm & \od & \od & \od
          &   & & 6 & $5^3$\\
          & \tm & \tm & \tm &  \tm&  \tm& \tm & \od & \od & \od
          &  $\otimes$ & & 12 & $5^{(1,3)}$\\
          & \tm & \tm & \tm &  \tm&  \tm& \tm & \od & \tm & $\otimes$
          & $\otimes$  &$\otimes$ & 4 & $6^{(3,1)}$\\
          \hline
    \end{tabular}
    \caption{Exotic brane embeddings corresponding to mixed symmetry potentials found from Bianchi identities. Here \tm{} denotes world-volume directions,  \od{} denotes quadratic special cycle, $\otimes$ denotes cubic special cycle.}
    \label{tab:exotic7}
\end{table}
Again we observe that the (6,1)-potential transforming in the $\mathbf{45}$ of SL(5) interacts with only 26 states. As in the NS-NS
case described by magnetic potentials of DFT we do not see a certain set of exotic potentials, e.g. $A_{(7,1,1),\mathbf{1}}$ corresponding to the embedding of the $8^{(1,0)}$-brane when the cubic cycle is in the external space. This must correspond to the R-flux of the theory that is generated by either dependence on winding coordinates or by turning on polyvector potentials, of which we have neither.  
 
\section{Discussion}

In this paper we have applied the paradigm of the complete flux formulation of double field theory to the SL(5) exceptional field theory. In the O(10,10) formulation of the former one has all fields collected into generalized vielbein and the generalized dilaton. Acting by generalized Lie derivative along the generalized vielbein on these fields one introduces the notion of generalized fluxes and the action and all field equations of DFT can be written in terms of these. In contrast in exceptional field theories (in general and in the SL(5) theory in particular) one finds the external vielbein and various tensor fields in addition to the generalized vielbein and the notion of a flux becomes less evident and is no longer a simple consequence of generalized Lie derivative along the generalized vielbein. 

The list of all fluxes of the theory is given in \eqref{sl5fluxes}. This begins with field strengths for the 1-, 2- and 3-form constructed in the usual way: start with commutator of covariant derivatives  $\mc{D}_\mu$ and proceed with adding gauge potentials to keep the expressions covariant. The scalar flux $\mF_{ABC}{}^D$ (the embedding tensor) and the anholonomy coefficients $\mc{F}_{\m\n}{}^\r$ are derived in the same way as in double field theory using generalized Lie derivative and ordinary Lie derivative respectively. The remaining two $\mc{F}^{(E)}_\m{}_A{}^B$ and $\mc{F}_MN{}_\bm{}^\bn$ are derived using the simple idea the fluxes must be composed of derivatives $\mc{D}_\m$ or $\dt_{MN}$ of the fundamental fields and the previous expressions do not include the external derivative of the generalized vielbein and the internal derivative of the external vielbein. One may notice that $\mc{F}_{\m\n}{}^\r$ already contains $\dt_{MN}$ acting on the external vielbein hidden inside $\mc{D_\m}$, however we find that $\mc{F}_{MN \bm}{}^\bn$ actually enters the Lagrangian and is an independent flux.

For the derived fluxes we find Bianchi identities both in curved (mixed) and all flat indices and construct full flux Lagrangian of the theory. The identities allow to define magnetic potentials for those of the SL(5) ExFT and to identify the corresponding (wrapped) branes, which we list in Section \ref{sec:exotic}. Among the found states we find those listed in \cite{Kleinschmidt:2011vu} and additional branes corresponding to wrappings of exotic membranes with at least one special direction left in the external space.

The presented results can be used and extended in several ways. The most obvious would be to derive flux formulation for other exceptional field theories. More fascinating directions of further research are related to the formalism of polyvector deformations \cite{Gubarev:2020ydf}. In the case in question these are defined as such SL(5) transformation generated by a tri-Killing that map a solution of 11D supergravity equations to a solution. Since field equations can be written in terms of generalized fluxes it is natural to formulate this condition as invariance of fluxes that allows to deal with first order equations rather than second order equations of motion. In \cite{Bakhmatov:2020kul,Gubarev:2020ydf} a special truncation has been imposed such that ExFT looks pretty much like DFT and only the scalar flux $F_{ABC}{}^D$ and a trace part of $\mc{F}_{MN \bm}{}^\bn$ are non-vanishing. This sensibly restricts the set of solutions that fit the truncation ansatz and does not allow e.g. the majority of the solutions found in \cite{Wulff:2016vqy}. Full flux formulation and the full list of fluxes presented here on allows to apply the above logic to any background. For that one has to derive conditions on the tri-Killing deformation ansatz sufficient for all fluxes to stay invariant. 

Another direction is related to employment of the derived Bianchi identities to extended the construction of \cite{Bakhmatov:2022lin} generalizing the same truncation of the 11D supergravity to the full theory. In this approach we first notice that non-unimodular generalized Yang-Baxter deformations transform generalized fluxes in a certain controllable way. To write ExFT equations for such transformed fluxes back in terms of space-time fields one has to impose Bianchi identities. These give certain conditions on the non-unimodularity parameter. We hope to report progress in these directions in the nearest future.

\section*{Acknowledgments}  This work has been supported by Russian Science Foundation grant RSCF-20-72-10144 and in part by the Foundation for the Advancement of Theoretical Physics and Mathematics “BASIS”, grant No 21-1-2-3-1.

\appendix 

\section{Notations and conventions}\label{NotesRefs}

In this paper, we use the following conventions for indices
\begin{equation}
    \begin{aligned}
       &\hat{\mu}, \hat{\nu},\ldots = 1 \dots 11&& \mbox{eleven directions, curved}; \\
       &\hat{\alpha}, \hat{\beta},\ldots = 1 \dots 11&& \mbox{eleven directions, flat}; \\       
       &\mu, \nu, \rho, \ldots = 1,\dots,7 && \mbox{external  directions (curved) of SL(5) ExFT}; \\
       &\bar{\mu}, \bar{\nu}, \bar{\rho}, \ldots 1,\dots,7 && \mbox{external   directions (flat) of SL(5) ExFT}; \\  
       &k, l, m, n,  \ldots = 1,\dots,4 && \mbox{internal directions (curved)}; \\       
       & a, b, c, d, \ldots = 1,\dots,4 && \mbox{internal directions (flat)}; \\  
       & \mM, \mN, \mK, \mL, \ldots = 1,\dots,10 && \mbox{SL(5) ExFT generalized space (curved)};  \\
       & \mA, \mB, \mC, \mD, \ldots = 1,\dots,10 && \mbox{SL(5) ExFT generalized space (flat)};\\
       & M, N, K, L, \ldots = 1,\dots, 5 && \mbox{fundamental SL(5) (curved)};  \\
       & A, B, C, D, \ldots = 1,\dots, 5 && \mbox{fundamental SL(5) (flat)}.
    \end{aligned}
\end{equation}

Generalized space of SL(5) ExFT is parametrized by coordinates $\mathbb{X}^{\mM}$. In terms of fundamental $\bf{5}$ indices of SL(5) they take form $\mathbb{X}^{MN}=-\mathbb{X}^{NM}$. The transition from $\bf{10}$ to antisymmetric pair of $\bf{5}$ is performed as
\begin{equation}
\begin{aligned}[]
T^{\mM} & \rightarrow T^{MN},\qquad\text{any tensor} , \\
U^{\mM} V_{\mM} & \rightarrow \frac12 U^{MN} V_{MN},\\
\delta^{\mM}_{\mN} & \rightarrow 2 \delta^{MN}_{KL},\qquad \text{only Kronecker}.
\end{aligned}
\end{equation}
The additional factor of two stands is needed to ensure $\delta^{\mM}_{\mM}=\frac12(2\delta^{MN}_{MN})=\delta^{MN}_{MN}=10$.

Epsilon tensors $\epsilon^{M N P Q R}$ and $\epsilon^{A B C D E}$ are absolutely antisymmetric tensors taking values $\pm1$ and are related as
\begin{equation}\label{relationepsilonsapp}
    \epsilon^{M N P Q R} E^{A}{}_{M} E^{B}{}_{N} E^{C}{}_{P} E^{D}{}_{Q} E^{E}{}_{R} = E \, \epsilon^{A B C D E} = e_{(7)}^{- \frac{5}{14}} \epsilon^{A B C D E}.
\end{equation}

\section{SL(5) algebra}\label{SL5algebra}
Here we give a brief review of the SL(5) algebra \cite{Musaev:2015ces}.
SL(5) generators in the fundamental $\bf{5}$ and the representation $\bf{10}$ read
\begin{equation}\label{sl5generators}
    \begin{split}
        (t^{I}{}_{J})^{M}{}_{N} \ &= \ \delta_{J}^{M} \delta_{N}^{I} - \frac15 \delta_{N}^{M}\delta_{J}^{I}, \\
        (t^{I}{}_{J})^{MN}{}_{KL} \ &= \ 4 (t^{I}{}_{J})^{[M}{}_{[K} \delta_{L]}^{N]}.
    \end{split}
\end{equation}
They satisfy the standard commutation relations
\begin{equation}\label{sl5generatorscommut}
        [t^{M}{}_{N},t^{K}{}_{L}] = \delta_{L}^{M} t^{K}{}_{N} - \delta_{N}^{K} t^{M}{}_{L},
\end{equation}
and the matrix product is defined as
\begin{equation}\label{sl5generatorstt}
        (t_{J}^{I}t_{L}^{K})^{\mM}\,_{\mN} = (t_{J}^{I})^{\mM}\,_{\mK}(t_{L}^{K})^{\mK}\,_{\mN} = \frac12 (t_{J}^{I})^{\mM}\,_{PQ}(t_{L}^{K})^{PQ}\,_{\mN}.
\end{equation}
Projector to the adjoint representation of SL(5) in the $\mathbf{10}$ irrep is given by
\begin{equation}\label{sl5project1}
        \mathbb{P}^{\mM}\,_{\mN}\,^{\mK}\,_{\mL} = \frac13 \kappa_{I_1 I_2}{}^{J_1 J_2} (t^{I_1}{}_{J_1})^{\mM}\,_{\mN}(t^{I_2}{}_{J_2})^{\mK}\,_{\mL} ,
\end{equation}
where the Killing form reads
\begin{equation}\label{CartanSL5form}
\kappa_{I_1 I_2}{}^{J_1 J_2} = \frac1{12} (t_{I_1}{}^{J_1})^{MN}{}_{KL} (t_{I_2}{}^{J_2})^{KL}{}_{MN} = \delta_{I_1}^{J_2} \delta_{I_2}^{J_1} - \frac15 \delta_{I_1}^{J_1} \delta_{I_1}^{J_1}.
\end{equation}
The projector satisfies the standard relation
\begin{equation}\label{sl5project2}
        \mathbb{P}^{\mM}\,_{\mN}\,^{\mK}\,_{\mL} \mathbb{P}^{\mL}\,_{\mK}\,^{\mP}\,_{\mQ} = \frac14 \mathbb{P}^{\mM}\,_{\mN}\,^{KL}\,_{IJ} \mathbb{P}^{IJ}\,_{KL}\,^{\mP}\,_{\mQ} = \mathbb{P}^{\mM}\,_{\mN}\,^{\mP}\,_{\mQ},
\end{equation}
and we also have $ \mathbb{P}^{\mM}\,_{\mN}\,^{\mN}\,_{\mM} = \frac14 \mathbb{P}^{M N}\,_{KL}\,^{KL}\,_{MN} = 24 = dim(adj)$. 
The $Y$-tensor of SL(5) ExFT and the projector are related as follows \cite{Berman:2012vc}
\begin{equation}\label{sl5Yproject}
        \epsilon^{M \mM \mK} \epsilon_{M \mN \mL} = Y^{\mM \mK}\,_{\mN \mL} = - 3 \mathbb{P}^{\mM}\,_{\mN}\,^{\mK}\,_{\mL} + \frac15 \delta^{\mM}_{\mN} \delta^{\mK}_{\mL} + \delta^{\mM}_{\mL} \delta^{\mK}_{\mN},
\end{equation}
where $\epsilon^{M \mM \mK}$ is the totally antisymmetric tensor of SL(5) $\epsilon^{M N K L P}$. We will also need the relation
\begin{equation}\label{sl5Yproject2}
        \epsilon^{TMNKL} \epsilon_{TPQRS} = - 3 \mathbb{P}^{MN}\,_{PQ}\,^{KL}\,_{RS} + \frac45 \delta^{MN}_{PQ} \delta^{KL}_{RS} + 4 \delta^{MN}_{RS} \delta^{KL}_{PQ}.
\end{equation}

\section{Generalized diffeomorphisms and weights}\label{gendiffandweight}

Consider transformations under generalized diffeomorphisms of a scalar of weight zero 
\begin{equation}
    \d_{\L} \phi = \frac{1}{2} \L^{M N} \dt_{M N} \phi.
\end{equation}
For its derivative we have the following transformation law
\begin{equation}
\begin{aligned}\label{transformofscalarderivative}
    \d_{\L} \dt_{K L} \phi &=  \dt_{K L} \d_{\L} \phi = \frac{1}{2} \dt_{K L} (\L^{M N} \dt_{M N} \phi) = \frac{1}{2} \L^{M N} \dt_{K L} \dt_{M N} \phi + \frac{1}{2} \dt_{M N} \phi \, \dt_{K L} \L^{M N} \\
    &= \frac{1}{2} \L^{M N} \dt_{M N} \dt_{K L} \phi + \dt_{K M} \phi \, \dt_{L N} \L^{M N} + \dt_{M L} \phi \, \dt_{K N} \L^{M N} - \frac{1}{2} \dt_{K L} \phi \, \dt_{M N} \L^{M N}
\end{aligned}
\end{equation}
where in the last equality we used the identity $    \dt_{[M N} \phi \, \dt_{K L]} \L^{M N} = 0 $, that follows from the section constraint (\ref{2cond}). Hence, one concludes that $\dt_{MN}\f$ is a generalized covector with weight $\l = - \frac15 $ and $\tilde{\lambda} = - 1$. As a result we see that derivative $\dt_{M N}$ adds certain weight. Note that if the scalar $\phi$ had a non-vanishing weight $\l_\f$, i.e. transformed as
\begin{equation}
    \d_{\L} \phi = \frac{1}{2} \L^{M N} \dt_{M N} \phi + \frac{\lambda_{\phi}}{2} \phi \, \dt_{M N} \L^{M N},
\end{equation}
its derivative $\dt_{K L} \phi$ would not transform as generalized covector.

Consider now a generalized vector $V^{M N}$ of weight $\l[V] = \frac65 $ ($\tilde{\lambda} = 2$) whose transformation reads
\begin{equation}
    \d_{\L}V^{M N} = \frac{1}{2} \L^{K L} {\dt}_{K L}{{V}^{M N}}\,  - 2 V^{L [N} {\dt}_{L K}{{\L}^{M] K}}\, + V^{M N} {\partial}_{K L}{{\L}^{K L}}.
\end{equation}
Let us now show that its derivative $\dt_{M N} V^{M N}$ transforms covariantly
\begin{equation}
\begin{aligned}
    \d_{\L} \dt_{M N} V^{M N} =& \, \frac{1}{2} \L^{K L} {\dt}_{K L}\dt_{M N}{{V}^{M N}}\, + \left( \frac{1}{2} \dt_{M N} \L^{K L} {\dt}_{K L}{{V}^{M N}}\, - 2 \dt_{M N}V^{L N} {\dt}_{L K}{{\L}^{M K}}\, \right. \\
    & \left.+ \dt_{M N} V^{M N} {\partial}_{K L}{{\L}^{K L}}\right) + \left(- 2 V^{L N} \dt_{M N} {\dt}_{L K}{{\L}^{M K}}\, + V^{M N} \dt_{M N} {\partial}_{K L}{{\L}^{K L}}\right), \\
    =& \, \frac{1}{2} \L^{K L} {\dt}_{K L}\dt_{M N}{{V}^{M N}}\, + \frac12 \dt_{M N} V^{M N} {\partial}_{K L}{{\L}^{K L}}.
\end{aligned}
\end{equation}
Where we have used the section condition rewritten as
\begin{equation}
    3 \dt_{[M N} \L^{K L} {\dt}_{K L]}{{V}^{M N}} + \frac12 \dt_{M N} V^{M N} {\partial}_{K L}{{\L}^{K L}} = \frac12 \dt_{M N} V^{M N} {\partial}_{K L}{{\L}^{K L}}
\end{equation}
to simplify terms in the first parentheses and the identity  $3 V^{M N} \dt_{[M N} {\partial}_{K L]}{{\L}^{K L}} = 0$ for the terms in the second parentheses.

\section{Derivation of Bianchi identities in flat indices}
\label{app:BIflat}

\textbf{Bianchi identity 1.}

Let us first define symbols $f_{AB,C}{}^D = E^D{}_M\dt_{AB}E_C{}^M$, where again  $\dt_{AB}=E_A{}^M E_B{}^N\dt_{MN}$, and we transform flat SL(5) indices to curved by generalized vielbein in SL(5)$ \times \mathbb{R}^{+}$. It can be rewritten as . Taking into account that $E\in \bf 24\oplus 1$ of the group SL(5) such defined tensor can be decomposed into irreducible representations as
\begin{equation}
f_{AB,C}{}^D \in \bf 10\otimes (24\oplus 1) = 10\oplus 10\oplus 15 \oplus \overline{40} \oplus 175.
\end{equation}
In terms of component tensor the decomposition can be written as follows
\begin{equation}\label{fabcdhelpdecomp}
f_{A B, C}{}^{D} = \frac{2}{9}f_{A B}\delta^{D}{}_{C}+\frac{1}{9}f_{C [A}\delta^{D}{}_{B]}-\frac{1}{9}q_{A B}\delta^{D}_{C}
-\frac{5}{9}q_{C [A}\delta^{D}{}_{B]} -\frac{1}{2}Y_{C [A}\delta^{D}{}_{B]}+ Z_{A B C}{}^D + \X_{A B C}{}^D,
\end{equation}
where
\begin{equation}
\begin{aligned}
f_{A B}&=f_{AB,C}{}^C && \in \bf 10,\\
q_{A B}&= f_{C[B,A]}{}^C && \in \bf 10, \\
Y_{A B}&=f_{C(A,B)}{}^C && \in \bf 15, \\
Z_{A B C}{}^{D} &= -\frac{4}{3}\epsilon_{A B C E F} Z^{E F D} &&\in \bf \overline{40}, \\
\X_{A B C}{}^D & && \in \bf 175.
\end{aligned}
\end{equation}
For the $\bf \overline{40}$ one has $Z^{[A B C]}\equiv 0$, and for the $\bf 175$ one has $\X_{[A B C]}{}^D\equiv0$, $\X_{A B,C}{}^{A} \equiv 0$, $\X_{A B,C}{}^{A}\equiv 0$. Note that from the two irreps $\bf 10$ only the combination $\q_{A B}=\frac{1}{10}(f_{A B} + q_{A B})$ survives when considering generalized Lie derivative as well as the full action. 
\begin{equation}
\begin{aligned}[]
\d_{E_{A B}} E_{C}{}^{M} &= \mF_{A B C}{}^D E_{D}{}^{M},\\
\mF_{A B C}{}^D & \in \bf 10 \oplus 15 \oplus \overline{40}.
\end{aligned}
\end{equation}

\begin{equation}
\begin{aligned}\label{BI1start}
    3 \, \DD_{[\m}(e^{\bn}{}_{\n} e^{\br}{}_{\r]} E_{A}{}^{M} E_{B}{}^{N} \mF_{\bn \br}{}^{A B}) & \stackrel{\text{№}1}{=} - \frac1{16} \, \epsilon^{M N P Q R} \dt_{P Q} (e^{\bm}{}_{[\m} e^{\bn}{}_{\n} e^{\br}{}_{\r]} e_{(7)}^{- \frac37} E^{C}{}_{R} E^{-1} \mF_{\bm \bn \br C} e_{(7)}^{\frac1{14}}),\\
\end{aligned}
\end{equation}
using (\ref{sl5fluxes}) the left hand side can be rewritten as
\begin{equation}
    \begin{aligned}
         3 \, \DD_{[\m}(e^{\bn}{}_{\n} e^{\br}{}_{\r]} E_{A}{}^{M} E_{B}{}^{N} \mF_{\bn \br}{}^{A B}) = &\  3 \, E_{A}{}^{M} E_{B}{}^{N} e^{\bn}{}_{[\n} e^{\br}{}_{\r} \DD_{\m]}(\mF_{\bn \br}{}^{A B}) + 3 \, \mF_{[\m \n|}{}^{\bl} \mF_{\bl | \r]}{}^{M N} \\
         &+ 6 \, \mF^{(E)}{}_{[\m| C}{}^{[M} \mF_{|\n \r]}{}^{N] C},
    \end{aligned}
\end{equation}
and similarly the right hand side
\begin{multline}
    - \frac1{16} \, \epsilon^{M N P Q R} \dt_{P Q} (e^{\bm}{}_{[\m} e^{\bn}{}_{\n} e^{\br}{}_{\r]} e_{(7)}^{- \frac37} E^{C}{}_{R} E^{-1} \mF_{\bm \bn \br C} e_{(7)}^{\frac1{14}}) =\\
    - \frac1{16} e_{(7)}^{-\frac1{14}} \, \epsilon^{M N P Q R} E^{E}{}_{R} e^{\bm}{}_{[\m} e^{\bn}{}_{\n} e^{\br}{}_{\r]} \dt_{P Q}(e_{(7)}^{\frac1{14}} \mF_{\bm \bn \br E}) - \frac3{16} \, \epsilon^{M N P Q R} \mF_{P Q [\m|}{}^{\bl} \mF_{\bl | \n \r] R} \\
    - \frac1{16} \, e_{(7)}^{- \frac{5}{14}} \epsilon^{M N P Q R} \mF_{\m \n \r C} \dt_{P Q} ( E^{C}{}_{R} E^{-1} ).
\end{multline}
The last term here can be worked out using \eqref{fabcdhelpdecomp}
\begin{multline}
    - \frac1{16} \, e_{(7)}^{- \frac{5}{14}} \epsilon^{M N P Q R} \mF_{\m \n \r C} \dt_{P Q} ( E^{C}{}_{R} E^{-1} ) = - \frac1{16} \, e_{(7)}^{- \frac{5}{14}} \epsilon^{M N P Q R} \mF_{\m \n \r C} (- f_{PQ,R}{}^{C} + f_{PQ,D}{}^{D} E^{C}{}_{R}) = \\
    - \frac1{16} \, e_{(7)}^{- \frac{5}{14}} \epsilon^{M N P Q R} \mF_{\m \n \r C} \left(\underbrace{\frac23 f_{P Q} \delta_{R}{}^{C} + \frac23 q_{P Q} \delta_{R}{}^{C}}_{\frac{20}3 \theta_{EF} \delta_{D}{}^{C}} - Z_{P Q R}{}^{C} \right).
\end{multline}
Using above results and definition (\ref{relationepsilons}) we obtain the first Bianchi identity \eqref{BIf1}.

\textbf{Bianchi identity 2.}

Using (\ref{sl5fluxes})
\begin{equation}
    4 \, \DD_{[\m}(e^{\bn}{}_{\n} e^{\br}{}_{\r} e^{\bs}{}_{\s]} E^{A}{}_{M} \mF_{\bn \br \bs A}) = 6 \, \mF_{[\m \n|}{}^{\bl} \mF_{\bl | \r \s] M} + 4 \, \mF^{(E)}{}_{[\m |M}{}^{C} \mF_{|\n \r \s] C} + 4 \, E^{A}{}_{M} e^{\bn}{}_{[\n} e^{\br}{}_{\r} e^{\bs}{}_{\s} \DD_{\m]} \mF_{\bn \br \bs A},
\end{equation}
and
\begin{multline}
    \dt_{M N}( e^{\bm}{}_{\m} e^{\bn}{}_{\n} e^{\br}{}_{\r} e^{\bs}{}_{\s} e_{(7)}^{-\frac47} E_{C}{}^{N} E^{-1} \mF_{\bm \bn \br \bs}{}^{C} e_{(7)}^{\frac3{14}}) = e^{\bm}{}_{\m} e^{\bn}{}_{\n} e^{\br}{}_{\r} e^{\bs}{}_{\s} e_{(7)}^{-\frac3{14}} E_{C}{}^{N} \dt_{M N}( \mF_{\bm \bn \br \bs}{}^{C} e_{(7)}^{\frac3{14}})\\
    + 4 \, \mF_{M N [\m|}{}^{\bl} \mF_{\bl | \n \r \s]}{}^{N} + \mF_{\m \n \r \s}{}^{C} E \, \dt_{M N}( E_{C}{}^{N} E^{-1} ),
\end{multline}
the last term here due to (\ref{TYZfluxes}) is
\begin{equation}
    E \dt_{M N}( E_{C}{}^{N} E^{-1} ) =  E^{A}{}_{M} (10 \theta_{A B} - Y_{A B}) = E^{A}{}_{M} \left(2 \mF_{A B D}{}^{D} - \frac12 \mF_{D (A B)}{}^{D}\right).
\end{equation}
Using above in the identities \eqref{BI2} we obtain \eqref{BIf2}.

\textbf{Bianchi identity 3.}

To find the third identities in flat indices we need to work out the following
\begin{multline}
    - \frac3{10} \, e_{(7)}^{-1} \dt_{M N}(e_{(7)}^{\frac{5}{7}} E_{A}{}^{M} E_{B}{}^{N} \mF_{[\bm \bn}{}^{A B} \delta_{\bl]}{}^{\br} e_{(7)}^{\frac{2}{7}}) = - \frac3{10} \, e_{(7)}^{-\frac{2}{7}} \dt_{A B}(e_{(7)}^{\frac{2}{7}} \mF_{[\bm \bn}{}^{A B}) \delta_{\bl]}{}^{\br}\\
    - \frac3{10} \, E^{2}  \mF_{[\bm \bn}{}^{A B} \delta_{\bl]}{}^{\br} \dt_{M N}(E^{-2} E_{A}{}^{M} E_{B}{}^{N}),
\end{multline}
where the last term can be transformed as follows
\begin{multline}
    - \frac3{10} \, E^{2}  \mF_{[\bm \bn}{}^{A B} \delta_{\bl]}{}^{\br} \dt_{M N}(E^{-2} E_{A}{}^{M} E_{B}{}^{N}) = - \frac6{10} \,  \mF_{[\bm \bn}{}^{A B} \delta_{\bl]}{}^{\br} E \, E_{A}{}^{M} \dt_{M N}(E^{-1} E_{B}{}^{N}) = \\
    - 6 \,  \mF_{[\bm \bn}{}^{A B} \delta_{\bl]}{}^{\br} \theta_{A B} = - \frac{6}{5} \,  \mF_{[\bm \bn}{}^{A B} \delta_{\bl]}{}^{\br} \mF_{A B C}{}^{C}.
\end{multline}
Substituting this in \eqref{BI3} we obtain Bianchi identities \eqref{BIf3}.

\textbf{Bianchi identity 5.}

The Bianchi identities \eqref{BI5} can be rewritten in the following form
\begin{equation}
\begin{aligned}
    0  = & - 2\, {\DD}_{[\bm|}{{\mF}_{M N |\bn]}\,^{\bl}}  + \frac{4}{5}\, {\mF}_{M N [\bm}\,^{\br} {\delta}_{\bn]}\,^{\bl} {\mF^{(E)}}_{\br A}\,^{A} + 2\, {\mF}_{[\bm| \br}\,^{\bl} {\mF}_{M N |\bn]}\,^{\br} - {\mF}_{\bm \bn}\,^{\br} {\mF}_{M N \br}\,^{\bl}\\
    & + e_{(7)}^{-\frac17} {\partial}_{M N}(e_{(7)}^{\frac17} {{\mF}_{\bm \bn}\,^{\bl}})\,+ \frac{4}{5}\, e_{(7)}^{-\frac17} {\partial}_{M N}(e_{(7)}^{\frac17} {{\mF^{(E)}}_{[\bm| A}\,^{A}})\,  {\delta}_{|\bn]}\,^{\bl} .
\end{aligned}
\end{equation}
To rewrite this in flat indices one has to turn curved indices of the flux under the derivative in the first term into flat ones:
\begin{multline}
    2\, {\DD}_{[\bm|}(E^{A}{}_{M} E^{B}{}_{N} {{\mF}_{A B |\bn]}\,^{\bl}}) = 2\, E^{A}{}_{M} E^{B}{}_{N} {\DD}_{[\bm|}({{\mF}_{A B |\bn]}\,^{\bl}}) + 4\, {{\mF}_{A B [\bn}\,^{\bl}} {\DD}_{\bm]}(E^{[B}{}_{N}) E^{A]}{}_{M} \\
    = 2\, E^{A}{}_{M} E^{B}{}_{N} {\DD}_{[\bm|}({{\mF}_{A B |\bn]}\,^{\bl}}) + 4\, {{\mF}_{A B [\bn}\,^{\bl}} \mF^{E}_{\bm] C}{}^{[B} E^{A]}{}_{M} E^{C}{}_{N}.
\end{multline}
Using this we obtain the deisired identities \eqref{BIf5} with flat indices.

\textbf{Bianchi identity 6*.}

To write the combined Bianchi identities (\ref{BI6star}) in flat indices we start with
\begin{equation}
    2 {\DD}_{[\mu}(e^{\bn}{}_{\n]} {{\mF^{(E)}}_{\bn A}\,^{B}}) = 2 e^{\bn}{}_{[\n} {\DD}_{\mu]}({{\mF^{(E)}}_{\bn] A}\,^{B}}) + \mF_{\m \n}{}^{\bn} {{\mF^{(E)}}_{\bn A}\,^{B}}
\end{equation}
and rewrite
\begin{equation}
    - {\partial}_{A C}(e_{(7)}^{-\frac{2}{7}} e^{\bm}{}_{\m} e^{\bn}{}_{\n} {{\mF}_{\bm \bn}\,^{B C}} e_{(7)}^{\frac{2}{7}}) = - e_{(7)}^{-\frac{2}{7}} e^{\bm}{}_{\m} e^{\bn}{}_{\n} {\partial}_{A C}({{\mF}_{\bm \bn}\,^{B C}} e_{(7)}^{\frac{2}{7}}) - {{\mF}_{\bm [\n|}\,^{B C}} \mF_{A C |\m]}{}^{\bm}.
\end{equation}
Using the above in (\ref{BI6star}) we arrive at \eqref{BIf6s}.

\textbf{Bianchi identity 7.}

The identities \eqref{BI7}  can be rewritten in the following form
\begin{equation}
\begin{aligned}
    \DD_{\m} \mF_{A B C}{}^{E} \stackrel{\text{№}7}{=} & \, - \frac{1}{2}\, {\partial}_{[A B|}{{\mF^{(E)}}_{\mu |C]}\,^{E}}\,  + {\delta}_{[A|}\,^{E} {\partial}_{C D}{{\mF^{(E)}}_{\mu |B]}\,^{D\, }}\,  - \frac{1}{2}\, {\delta}_{C}\,^{E} {\partial}_{[A| D}{{\mF^{(E)}}_{\mu |B]}\,^{D\, }}\,  \\
    & - 2 {\mF}_{[A| D\,  C}\,^{E} {\mF^{(E)}}_{\mu |B]}\,^{D\, } - {\mF}_{A B D\, }\,^{E} {\mF^{(E)}}_{\mu C}\,^{D\, } + {\mF}_{A B C}\,^{D\, } {\mF^{(E)}}_{\mu D\, }\,^{E}.
\end{aligned}
\end{equation}
We will also need the following identity 
\begin{equation}
    {\partial}_{A B}({{e_{(7)}^{-\frac{1}{7}} e^{\bm}{}_{\m} \mF^{(E)}}_{\bm C}\,^{E}} e_{(7)}^{\frac{1}{7}}) = {{e_{(7)}^{-\frac{1}{7}} e^{\bm}{}_{\m} {\partial}_{A B}(\mF^{(E)}}_{\bm C}\,^{E}} e_{(7)}^{\frac{1}{7}}) + \mF_{A B \m}{}^{\bm} {\mF^{(E)}}_{\bm |C}\,^{E}.
\end{equation}
Using the above in the \eqref{BI7} we obtain precisely \eqref{BIf7}.

\textbf{Bianchi identity 8.}

Here we start with rewriting terms in the  identities \eqref{BI8} as follows
\begin{multline}
    - \partial_{M N}{\mF_{K L \bm}{}^{\bn}} + \partial_{K L} \mF_{M N \bm}{}^{\bn} = - E^{A B}{}_{K L} \partial_{M N}{\mF_{A B \bm}{}^{\bn}} + E^{A B}{}_{M N} \partial_{K L} \mF_{A B \bm}{}^{\bn} + \\ 
    + E_{A B}{}^{P Q} \mF_{P Q \bm}{}^{\bn} ( - \partial_{M N} E^{A B}{}_{K L} + \partial_{K L} E^{A B}{}_{M N}).
\end{multline}
Terms in the second line can be written back in terms of generalized fluxes, using the fact that $\mF_{P Q \bm}{}^{\bn}$ is antisymmetric in $P Q$
\begin{equation}
\begin{aligned}
    E_{A B}{}^{P Q} \mF_{P Q \bm}{}^{\bn} ( - \partial_{M N} E^{A B}{}_{K L} + \partial_{K L} E^{A B}{}_{M N})= 2 \, \mF_{[K| Q \bm}{}^{\bn} f_{M N, |L]}{}^{Q} - 2 \, \mF_{[M| Q \bm}{}^{\bn} f_{K L, |N]}{}^{Q}.
\end{aligned}
\end{equation}
Finally using the decomposition (\ref{fabcdhelpdecomp}) for $f_{A B, C}{}^{D}$, properties of its components and the section constraint we obtain (see Cadabra calculation \texttt{FMNmunu} in \cite{sl5flux:git})
\begin{equation}
\begin{aligned}
    2 \, \mF_{[K| Q \bm}{}^{\bn} f_{M N, |L]}{}^{Q} - 2 \, \mF_{[M| Q \bm}{}^{\bn} f_{K L, |N]}{}^{Q} = \, 2 \, \mF_{[K| Q \bm}{}^{\bn} \mF_{M N, |L]}{}^{Q} - 2 \, \mF_{[M| Q \bm}{}^{\bn} \mF_{K L, |N]}{}^{Q}.
\end{aligned}
\end{equation}
As the result we have
\begin{multline}
    - \partial_{M N}{\mF_{K L \bm}{}^{\bn}} + \partial_{K L} \mF_{M N \bm}{}^{\bn} = - E^{A B}{}_{K L} \partial_{M N}{\mF_{A B \bm}{}^{\bn}} + E^{A B}{}_{M N} \partial_{K L} \mF_{A B \bm}{}^{\bn} + \\ 
    + 2 \, \mF_{[K| Q \bm}{}^{\bn} \mF_{M N, |L]}{}^{Q} - 2 \, \mF_{[M| Q \bm}{}^{\bn} \mF_{K L, |N]}{}^{Q},
\end{multline}
that after substitution into \eqref{BI8} gives us \eqref{BI8}.

\providecommand{\href}[2]{#2}\begingroup\raggedright\endgroup

\end{document}